\documentclass[twocolumn,aps,showpacs,preprintnumbers,eqsecnum,amsfonts,amsmath,amssymb]{revtex4}

\usepackage{graphicx} 
\usepackage{dcolumn}  
\usepackage{bm}       
\usepackage{latexsym} 
\newcommand{\smR}{{\scriptscriptstyle R}}
\newcommand{\smC}{{\scriptscriptstyle C}}
\newcommand{\smS}{{\scriptscriptstyle S}}
\newcommand{\smI}{{\scriptscriptstyle I}}
\newcommand{\smH}{{\scriptscriptstyle H}}
\newcommand{\smV}{{\scriptscriptstyle V}}
\newcommand{\smB}{{\scriptscriptstyle B}}
\newcommand{\smD}{{\scriptscriptstyle D}}

\newcommand{\smGT}{{\scriptscriptstyle \,>}}
\newcommand{\smLT}{{\scriptscriptstyle <}}
\newcommand{\smDelQ}{{\scriptscriptstyle \Delta Q}}
\newcommand{\smPerp}{{\scriptscriptstyle \perp }}
\newcommand{\smParallel}{{\raise0pt\hbox{\kern1.5pt\rule{.2pt}{6pt}\kern1.5pt\rule{.2pt}{6pt}}}}

\begin{document}

\preprint{\hfill LA-UR-07-2154~~Phys. Rev. E {\bf 76} (2007) 066404}
%

\title{
 Temperature Equilibration Rate with Fermi-Dirac Statistics
}
\author{Lowell S. Brown}
\author{Robert L. \surname{Singleton Jr.}}
\affiliation{
     Los Alamos National Laboratory\\
     Los Alamos, New Mexico 87545, USA
}
\date{20 November 2007}
\begin{abstract}
\noindent
We calculate analytically the electron-ion temperature equilibration
rate in a fully ionized, weakly to moderately coupled plasma, using an
{\em exact} treatment of the Fermi-Dirac electrons. The temperature is
sufficiently high so that the quantum-mechanical Born approximation to
the scattering is valid. It should be emphasized that we do not build
a {\em model} of the energy exchange mechanism, but rather, we perform
a systematic first principles {\em calculation} of the energy
exchange.  At the heart of this calculation lies the method of
dimensional continuation, a technique that we borrow from quantum
field theory and use in a novel fashion to regulate the
kinetic equations in a {\em consistent} manner. We can then perform a 
systematic perturbation
expansion and thereby obtain a finite first-principles result to
leading and next-to-leading order. Unlike model building, this
systematic calculation yields an estimate of its own error and thus
prescribes its domain of applicability.  The calculational error is
small for a weakly to moderately coupled plasma, for which our result is
nearly exact. It should also be emphasized that our calculation
becomes unreliable for a strongly coupled plasma, where the
perturbative expansion that we employ breaks down, and one must then
utilize model building and computer simulations.  Besides
providing new and potentially useful results, we use this
calculation as an opportunity to explain the method of dimensional
continuation in a pedagogical fashion. Interestingly, in the regime
of relevance for many inertial confinement fusion experiments, 
the degeneracy corrections are
comparable in size to the subleading quantum correction below the Born
approximation. For consistency, we therefore present this subleading
quantum-to-classical transition correction in addition to the
degeneracy correction. 
\end{abstract}


\maketitle

\section{Introduction and Summary}

We shall calculate the thermal equilibration rate between Fermi-Dirac
electrons and Maxwell-Boltzmann ions in a hot, fully ionized
plasma. We shall do so {\em exactly} to leading and next-to-leading
order in the plasma number density, and to all orders in the electron
fugacity, thereby providing an essentially  exact result for weakly to
moderately coupled plasmas.  We shall work out this problem for two
reasons.  First, the result is new and is needed in some
applications. Second, our previous treatment in Ref.~\cite{bps} of the
plasma stopping power was performed in great generality, and the basic
idea behind the dimensional continuation method, a somewhat subtle
analytic tool that we employ, 
may have gotten lost in all the details. We shall use this
opportunity of a simpler and specific case to treat the method in a
pedagogical fashion and to explain it clearly~\cite{bpsex}.

Physical systems often contain disparate length or energy scales.  For
example, plasma physics involves hard collisions at short distances --
ultraviolet physics; and soft interactions at large distances
entailing collective effects -- infrared physics.  The resulting
interplay of short and long distances produces the familiar Coulomb
logarithm. For the electron-ion temperature equilibration rate, and
for other such processes involving disparate scales, it is rather easy to
calculate the leading contribution, namely the overall factor in front
of this logarithm. Although the order of magnitude of this leading order 
term can usually be obtained from simple dimensional analysis alone,
calculating the additional dimensionless factor inside the logarithm,
the sub-leading term, is quite difficult.

A new method~\cite{lfirst} employing dimensional
continuation has been introduced to deal with such problems, a
method that makes the computation of the sub-leading as well as
the leading contributions tractable.  This method is based on
tested principles of quantum field theory constructed over the
last fifty years, and it has been used successfully to calculate
well measured phenomena such as the Lamb shift~\cite{lfirst},
often with much more ease than traditional methods. Most recently,
the method has been exploited in Ref.~\cite{bps} by Brown,
Preston, and Singleton (BPS) to provide an extensive treatment of
the charged particle stopping power in a plasma, the energy loss
per unit distance $dE/dx$ of the charged projectile.  One of the
topics treated in BPS was the rate at which elections and ions in
a spatially homogeneous plasma, starting with different
temperatures, come into thermal equilibrium at a common
temperature. Here we shall extend this work to include the case
in which the electron fugacity is sufficiently large that the
electrons must be treated with a degenerate Fermi-Dirac
distribution. This is the case in which Pauli blocking becomes
of some importance.

The degeneracy effects that we treat here come into play as the plasma
temperature is lowered. We shall calculate the rate for the general
case in which the electrons are described by a Fermi-Dirac
distribution, with no approximations being made to this
distribution. That is, we shall perform the calculation exactly to all
orders in the electron fugacity
\begin{equation}
  z_e = e^{\beta_e \, \mu_e} \,,
\end{equation}
where $\beta_e = 1 / T_e$ is the reciprocal of the electron
temperature and $\mu_e$ is the electron chemical potential. Note that
we shall always measure temperature in
energy units so that $\mu_e$ does indeed have the correct units of 
energy. We shall assume that the plasma is at most
moderately coupled, which often implies that the degeneracy
corrections are not large. Nonetheless, we shall work out the 
general case since
this is just as easy as treating the case of only mild degeneracy, and
the general case may prove to have some application.

A plasma is seldom formed in thermal equilibrium; for example, a laser 
preferentially heats the light electrons rather than the heavy ions.  
While a non-equilibrium plasma will of course eventually thermalize, it 
does so in several stages. First, 
the electrons rapidly equilibrate among themselves to a
common temperature $T_e$.  Then, somewhat less rapidly, all the
various ions equilibrate to a common temperature $T_\smI$. 
Finally, the electrons and ions begin the process of thermal
equilibration, with the electrons delivering their energy density to
the ions at a rate
\begin{eqnarray}
  \frac{d{\cal E}_{e \smI}}{dt}
  =   
  -\,{\cal C}_{e \smI} \left( T_e - T_\smI \right) \ .
\label{dedteeii}
\end{eqnarray}

We shall calculate the rate coefficient ${\cal C}_{e\smI}$ in 
the following sections. However, it is useful to
examine this result now, the rate coefficient in the form
(\ref{donerdeal}), which we display below for convenience. 
This expression for ${\cal C}_{e\smI}$ is exact to all
orders in the electron fugacity, but valid only in a temperature
regime in which the short-distance scattering is described by the
quantum-mechanical first Born approximation: 
\begin{eqnarray}
&&  {\cal C}_{e \smI} 
  = 
  \frac{\omega_\smI^2}{2\pi}\,\sqrt{\frac{\beta_e m_e}{2\pi}} \,
  \left( \beta_e e^2\,\frac{2\,e^{\beta_e\mu_e}}{\lambda_e^3} \, 
  \right)
   \Bigg\{
  \frac{\ln\Lambda}{ \exp\!\left\{ \beta_e  \mu_e \right\}  + 1} \, 
\nonumber\\[5pt] 
&& \hskip2cm 
  +\, \frac{1}{2}
  \sum_{l=1}^\infty (-1)^{l+1}  \ln\{l+1\}\,  e^{l\beta_e\mu_e}\Bigg\} \ ,
\label{pauli}
\end{eqnarray}
where
\begin{equation}
\ln \Lambda = \frac{1}{2}\,
  \left[ \ln\!\left\{ \frac{16\pi}{\kappa_e^2\lambda_e^2} \right\}
  -\gamma - 1  \right] \,.
\label{Lambda}
\end{equation}
Here $\omega_\smI^2$ is the sum of the squared ion plasma 
frequencies~$\omega_i^2$, 
$m_e$ is the electron mass, $\lambda_e$ is the electron thermal wave
length, and $\kappa^2_e$ is the electron contribution to the squared
Debye wave number, {\em including} electron degeneracy effects.  The precise
definition of these quantities is presented in Section~\ref{note}, but
we should note here that we employ rationalized Gaussian units, so
that the energy of two electrons of charge $e$ a distance $r$ apart is
given by $e^2 / 4\pi r$. The structure of the first line that appears
in our result (\ref{pauli}) agrees with the previous result of 
Brysk~\cite{B} when his Eq.~(35) is re-expressed in terms of our notation.
However, Brysk does not obtain the precise result (\ref{Lambda}) for $\ln
\Lambda$, but rather only an approximate, leading-log evaluation of
the usual form $\ln\{b_\text{max}/b_\text{min}\}$. Moreover,
Brysk~\cite{B} also does not obtain the second line of our result
(\ref{pauli}).  This second line does not contribute in the
Maxwell-Boltzmann limit of very small fugactiy $z_e$, but it does a
provide a significant first-order correction in $z_e$.

The previous work of BPS~\cite{bps} computed the exact temperature
equilibration rate for a weakly to moderately coupled non-degenerate
plasma. While this general result was
rather complicated, the high-temperature limit, in which the
short-distance Coulomb scattering is given by its first Born
approximation, has a rather simple form. The non-degenerate rate
coefficient of BPS can be obtained from Eq.~(\ref{pauli}) by taking
the small-fugacity limit $z_e \to 0$, which gives
\begin{eqnarray}
  {\cal C}_{e \smI}^\text{\,non-degen}
  &=& 
  \frac{\omega_\smI^2}{2\pi}\,\sqrt{\frac{\beta_e m_e}{2\pi}} \,
  \left( \beta_e\, e^2\, n_e
  \right)
  \ln \Lambda_0 \ ,
\label{nondegenrate}
\end{eqnarray}
where the logarithm $\ln\Lambda_0$ above is the non-degenerate limit
of Eq.~(\ref{Lambda}). The term 
\begin{eqnarray}
  \kappa_{e\, 0}^{2} \equiv \beta_e\, e^2\, n_e
\label{kappa0}
\end{eqnarray}
in parentheses follows from the well
known relation (\ref{napprox}) between number density and fugacity,
and it is just the square of the non-degenerate form of the electron's
Debye wave number. The
non-degenerate limit of Eq.~(\ref{Lambda}) is accomplished by the
substitution $\kappa_e \to \kappa_{e\, 0}$, and expressing this result
in terms of the electron plasma frequency provides the from, 
\begin{equation}
  \ln \Lambda_0 = \frac{1}{2}\,
  \left[ \ln\!\left\{ \frac{8\, T_e^2}{\hbar^2 \omega_e^2} \right\}
  -\gamma - 1  \right] \,,
\label{LambdaBPS}
\end{equation}
where we have used the relation $\kappa_{e\, 0}^2\, \lambda_e^2 / 2\pi =
\hbar^2 \omega_e^2 / T_e^2 $.  
The rate coefficient of Eq's.~(\ref{nondegenrate}) and
(\ref{LambdaBPS}) is just that given previously by Eq.~(12.12) of
BPS~\cite{bps}, an expression that was also quoted in Eq.~(3.61) in
the introductory portion of that work.  There is, however, one
difference between Eq.~(\ref{LambdaBPS}) and the result (12.12) of
BPS. Namely, the correct term $-\gamma-1$ appearing in
Eq.~(\ref{LambdaBPS}) above was incorrectly
written as $-\gamma -2$ in Eq's.~(12.12) and (3.61) of BPS because of
a transcription error in passing between Eq's.~(12.43) and (12.44) of
BPS. We have taken the opportunity here to correct this mistake.

Equation~(\ref{LambdaBPS}) gives the precise definition of the
``Coulomb log'' (in the non-degenerate limit) for the temperature
equilibration process we have been discussing, including the correct
additional constant terms, namely the terms $\ln 8 - \gamma -1$. We
should, however, emphasize two points.  First of all, the ``Coulomb
log'' is by no means a universal quantity, but rather its precise form
is process dependent.  For example, the ``Coulomb log'' for electron
conductivity differs from the ``Coulomb log'' for the electron-ion
thermal relaxation rate given here~\cite{BCH}. Secondly, we should
emphasize that the ``Coulomb log'' for the relaxation rate does not
depend upon the ion temperature $T_\smI$ but only upon the electron
temperature $T_e$.  Some authors~\cite{BCH} incorrectly replace the
squared electron Debye wave number $\kappa_e^2$ by the fully screened
Debye wave number $\kappa_\smD^2 = \kappa_e^2 + \kappa_\smI^2$. This
incorrectly includes the ion contribution $\kappa_\smI^2$, and thus
introduces a spurious dependence on the ion temperature $T_\smI$. 

If the plasma temperature becomes very low, then the population of
bound states must be taken into account. In such a regime, our
assumption that the plasma is fully ionized and weakly coupled breaks
down, and our calculation is no longer reliable. Hence we shall
discuss and examine in detail only the limit in which the degeneracy
corrections are mild, the regime in which the temperature is not too
low and the electron fugacity is not too large. In this limit, only
the first-order fugacity contribution from the general result
(\ref{pauli}) need be retained, and therefore the second line in
Eq.~(\ref{pauli}), the term omitted by Brysk~\cite{B}, makes an 
essential contribution.  As the temperature
is lowered, however, the subleading contribution to the Born
approximation becomes comparable to the degeneracy correction, and it
too must also be accounted for. In Appendix~\ref{sec:scatcorr}, we
therefore extract this subleading correction in the transition region
between quantum and classical scattering from the general result given
in BPS. In the following, these two types of corrections will be
called the degeneracy correction and the (first) quantum-to-classical
transition correction. 

Working to leading order in the fugacity, we only need to expand
Eq.~(\ref{pauli}) to linear order in $z_e=e^{\beta_e \mu_e}$, a result
contained in Eq.~(\ref{agreed}).  The quantum-to-classical
transition correction is contained in Eq.~(\ref{firstcl}). Expressing
the fugacity correction in terms of the density according to
Eq.~(\ref{fugacity}), the rate coefficient ${\cal C}_{e \smI}$ reads
\begin{widetext}
\begin{eqnarray}
  {\cal C}_{e \smI} 
  &\simeq& 
  \frac{\omega_\smI^2}{2\pi}\,\sqrt{\frac{\beta_e m_e}{2\pi}}\,
  \Big( \beta_e \, e^2 n_e\Big)
  \Bigg\{\frac{1}{2} \,\left[ \ln\!\left\{\frac{8 T_e^2}
  {\hbar^2 \omega_e^2} \right\} - \gamma - 1  \right] 
\nonumber\\[5pt] && \qquad 
  +\, \frac{n_e \lambda_e^3}{2} \, \left[
  -\left( 1 - \frac{1}{2^{3/2}} \right) \,\frac{1}{2}
  \left[ \ln\!\left\{  \frac{8 T_e^2}{\hbar^2\omega_e^2}\right\}
   - \gamma - 1  \right] 
   + \left(\frac{1}{2}\,\ln 2  + \frac{1}{2^{5/2}} \right) \right] 
\nonumber\\[5pt] && \qquad
  -\, \frac{\epsilon_\smH}{T_e}\, {\sum}_i \, 
  \frac{Z_i^2 \, \omega_i^2}{\omega_\smI^2} \,
  \left[ \zeta(3)\left(
  \ln\!\left\{\frac{T_e}{Z_i^2\,\epsilon_\smH} \right\}
  -\gamma \right) -2\,\zeta'(3) \right]
\Bigg\} \,,
\label{answer}
\end{eqnarray}
\end{widetext}
where the numerical values of the zeta-function and its
derivative are 
\begin{eqnarray}
  \zeta(3) 
  =  
  \sum_{k=1}^\infty \, \frac{1}{k^3} = 1.20205 \cdots \,,
\end{eqnarray}
and
\begin{eqnarray}
  \zeta'(3) 
  =  
  -\sum_{k=1}^\infty \, \frac{1}{k^3} \, \ln k = -0.19812 \cdots \,.
\end{eqnarray}
We also write $Z_i$ as the ionic charges in units of the electron
charge~$e$.  Before examining expression (\ref{answer}) in detail, we
note that the first line is the leading rate coefficient calculated in
BPS~\cite{bps}, Eq.~(\ref{nondegenrate}) above; the second line
is the first degeneracy correction following from Eq.~(\ref{pauli}); 
and the third line is the first quantum-to-classical transition
calculated in Appendix~\ref{sec:scatcorr}. 

In the last line of Eq.~(\ref{answer}), the ratio $\epsilon_\smH/T_e$
describes the relative size of the correction, where
\begin{eqnarray}
  \epsilon_\smH 
  = 
  \left(\frac{e^2}{4\pi}\right)^2 \,\frac{m_e}{2 \, \hbar^2} 
  \simeq 13.6 \, {\rm eV}
\label{bind}
\end{eqnarray}
is the binding energy of the hydrogen atom. 
For some temperature and number density regimes
of interest, the second and third lines in Eq.~(\ref{answer}) become
comparable in size.  Hence, while our main thrust in this paper is
degeneracy corrections, we must also take into account this first
quantum-to-classical transition. 

It is conventional to write the Coulomb logarithm as 
$\ln\{b_\text{max}/b_\text{min}\}$, where $b_{\max}$ is a
Debye length long-distance cutoff, while $b_\text{min}$ is a
short-distance cutoff that, depending upon the circumstances, is 
either a classical distance of closest approach 
$b_\text{cl} \sim e^2 / T_e$ or a quantum wave length 
$b_\text{qm} \sim \lambda_e$. Often, an interpolation is
made~\cite{LP} between these two limits by writing
\begin{eqnarray} 
 b_\text{min}^2 &=& b_\text{cl}^2 + b_\text{qm}^2  
= b_\text{qm}^2 \left[ 1 + \frac{b_\text{cl}^2}{b_\text{qm}^2} \right] 
\nonumber\\
&\sim&  b_\text{qm}^2 \left[ 1 + \frac{\epsilon_\smH}{T_e} \right] \,. 
\end{eqnarray}   
Such an interpolation gives a first correction proportional to the
proper quantum expansion parameter $\epsilon_\smH/T_e$, but it fails
entirely to produce the proper logarithmic behavior
$(\epsilon_\smH/T_e) \ln\{\epsilon_\smH/T_e\}$
displayed in the last line of Eq.~(\ref{answer}). 

\begin{figure}
\includegraphics[scale=0.40]{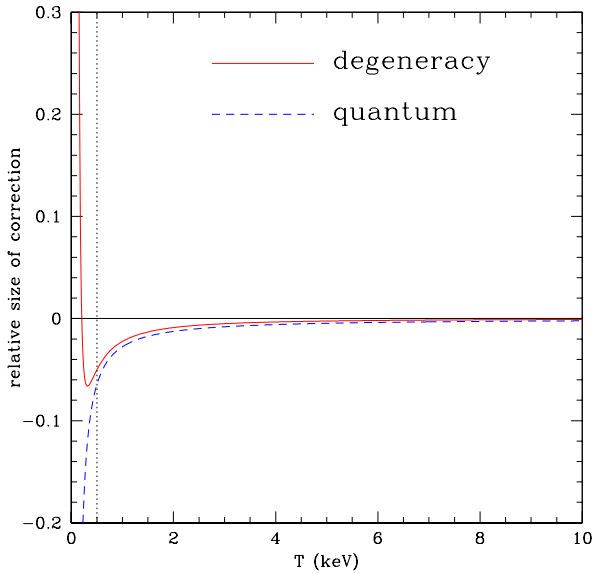}
\caption{
  The relative size of the degeneracy correction and the first
  classical-to-quantum correction as a function of temperature in
  keV. The plasma is equimolar deuterium-tritium at an electron number
  density $n_e=10^{25}\, {\rm cm}^{-3}$. These corrections are
  relative to the leading non-degenerate BPS rate: the degeneracy
  correction (the solid line) is the ratio of the second to the first
  line in Eq.~(\ref{pauli}), while the classical-to-quantum
  correction (the dashed line) is the ratio of the third to the
  first. The electron temperature runs between values $0.1\,{\rm keV}$
  and $10\,{\rm keV}$. Our calculation ceases to be valid at low
  temperatures, and this is indicated by the vertical dotted line.
}
\label{fig:ratio25}
\end{figure}
\begin{figure}[t]
\includegraphics[scale=0.40]{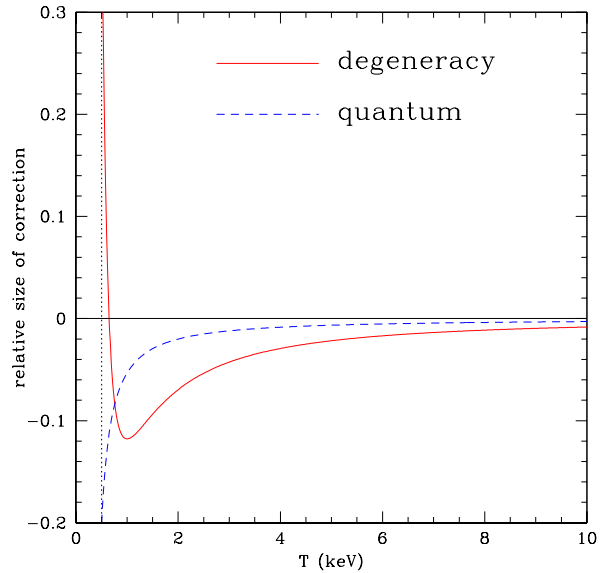}
\caption{
  Same as Fig.~\ref{fig:ratio25}, except the number density is
  \hbox{$n_e = 10^{26}\, {\rm cm}^{-3}$}. Again, the solid line is the
  degeneracy correction and the dashed line is the first
  quantum-to-classical correction.
}
\label{fig:ratio26}
\end{figure}

Figures~\ref{fig:ratio25} and \ref{fig:ratio26} illustrate the size of
the degeneracy and the first quantum-to-classical transition
corrections in an equimolar deuterium-tritium plasma, for electron
number densities of $n_e=10^{25}\,{\rm cm}^{-3}$ and
$n_e=10^{26}\,{\rm cm}^{-3}$, respectively. The solid curves denote
the size of the degeneracy corrections relative to the leading BPS
term, the ratio of the second to the first line of
Eq.~(\ref{answer}).  The dashed curves show the corresponding size of
the first quantum-to-classical transition, the ratio of the third to
the first line of Eq.~(\ref{answer}). Note that both corrections are
comparable in magnitude between these two electron number densities. 
We have plotted the
corrections for temperatures between $T_e=0.1\,{\rm keV}$ and
$T_e=10\,{\rm keV}$.  However, below about $0.5\,{\rm keV}$,
ionization and other strongly coupled plasma effects become important,
and our formalism breaks down. Also, at these lower temperatures, 
higher order fugacity terms in Eq.~(\ref{Lambda}) become important,
and this could change the low-temperature behavior of the solid
curves; therefore, one should trust Eq.~(\ref{answer}) 
only to the right of the vertical dotted line at $T_e=0.5\, {\rm keV}$. 
For densities below
$n_e=10^{25}\, {\rm cm}^{-3}$, the degeneracy correction is much
smaller than the quantum-to-classical correction. The situation is
reversed for densities greater than $n_e=10^{26}\,{\rm cm}^{-3}$,
where degeneracy effects dominate over the quantum-to-classical
corrections.  Interestingly, the density and temperature range in
which the degeneracy and the quantum-to-classical corrections are
comparable lies in the regime relevant for inertial confinement
fusion.

We turn now to present the details of our work. After establishing the
conventions and notation that we employ, we review and explain
in some detail the dimensional continuation method that forms the
basis of our calculation, and we relate this method to previous work that
utilized convergent kinetic equations. After establishing this
foundation, we then perform the calculation of the
degeneracy corrections to the rate at which the electrons and ions
come into thermal equilibrium. Appendix~\ref{sec:difun} contains the
details of the plasma dielectric function needed in the text, and
Appendix~\ref{sec:scatcorr} extracts the first correction in the
transition to classical scattering from a general formula given in 
BPS~\cite{bps}.

But before passing to these details, we would like to make a
comment.  There are a large number of very able
theoretical physicists who work on formal developments that have
no contact with experiment. In the chance that one of them may
encounter this paper, we would like to encourage them to once in
a while apply their skills to topics such as that treated here.
The field would profit from their work. In this regard, we quote
a penciled poem written in 1957 on note-paper embossed with the
header ``The White House, Washington'':

\vskip0.3cm 
\begin{tabular}{l}
{\em A  fact without a theory is like a ship without a sail,}\\[3pt]
{\em is like a boat without a rudder,}\\[3pt]
{\em is like a kite without a tail.}\\[6pt]
{\em A fact without a theory is like an inconsistent act.}\\[3pt]
{\em But if there is one thing worse in this confusing}\\
{\em  \hskip0.2cm universe,}\\[3pt]
{\em it is a theory without a fact.}
\end{tabular}

\section{\label{note} Conventions and Notation}

We will treat the ions with Maxwell-Boltzmann statistics
and the electrons with Fermi-Dirac statistics.  The thermal equilibrium 
form of an ion phase space density $  f_i({\bf p}_i) $ thus reads
\begin{eqnarray}
  f_i({\bf p}_i) = e^{-\beta_i\, (E_i({\bf p}_i) - \mu_i)} \,,
\label{defFiA}
\end{eqnarray}
while for electrons in thermal equilibrium,
\begin{eqnarray}
  f_e({\bf p}_e) = \frac{1}{e^{\beta_e\, (E_e({\bf p}_e) - \mu_e)} + 1} \,.
\label{defFeA}
\end{eqnarray}
Letting the index $b$ refer to either the ions $i$ 
or the electrons $e$, the kinetic energy is 
\begin{eqnarray}
  E_b({\bf p}_b) = \frac{p_b^2}{2 m_b} \ ,
\end{eqnarray}
and the inverse temperature and chemical potential are
\begin{eqnarray}
  \beta_b=1/T_b \hskip0.5cm \text{and}\hskip0.5cm \mu_b \ .
\end{eqnarray}
Since we shall work in an arbitrary number of dimensions $\nu$, each
species number density appears as
\begin{eqnarray}
  n_b 
  = 
  \mathfrak{g}_b\! \int\frac{d^\nu p_b}{(2\pi\hbar)^\nu}\,f_b({\bf p}_b) \ ,
\label{fbnorm}
\end{eqnarray}
where $\mathfrak{g}_b$ is spin-degeneracy factor. For electrons
$\mathfrak{g}_e=2$.  
We are using a notation for the distribution functions $f_b$ in which
the species index $b$ implicitly includes spin degrees of freedom.  It
is inconvenient, however, to use this notation for the number density
$n_b$. Measurements of the species density are usually insensitive to
spin degrees of freedom, and we shall therefore denote the number
density of the species (including all the spins) by $n_b$. This
accounts for the factor of $\mathfrak{g}_b$ in Eq.~(\ref{fbnorm}).
For the ions in thermal equilibrium, the integral
(\ref{fbnorm}) is a product of trivial Gaussian integrals, and so
\begin{eqnarray}
  n_i 
  = 
  \mathfrak{g}_i \, \lambda_i^{-\nu} \,  e^{\beta_i \, \mu_i} \,,
\label{MBn}
\end{eqnarray}
where we define the thermal wavelength for species $b$ as
\begin{eqnarray}
  \lambda_b
  = 
  \hbar \left( \frac{2\pi \, \beta_b}{m_b}\right)^{1/2} \ .
\label{deflambd}
\end{eqnarray}
For the electrons in thermal equilibrium, we first pass to hyper-spherical
coordinates to write
\begin{eqnarray}
  n_e  
  = 
  \mathfrak{g}_e \, \frac{\Omega_{\nu-1}}{(2\pi\hbar)^\nu}
  \int_0^\infty \!\! p^{\nu-1}\, dp \,
  \frac{1}{e^{\beta_e\, (E_e({\bf p}) - \mu_e)} + 1} \,.
\end{eqnarray}
Here $ \Omega_{\nu-1} $ is the area of a unit hyper-sphere in a space 
of $\nu$ dimensions that is evaluated in the next section with the result
\begin{eqnarray}
  \Omega_{\nu-1} =  \frac{2 \,\pi^{\nu/2}}{\Gamma\left(\nu/2\right)} \,.
\end{eqnarray}
Changing to dimensionless variables, 
\begin{eqnarray}
  x = \beta_e \, E_e({\bf p}) = \beta_e \, \frac{p^2}{2m_e} \ ,
\end{eqnarray}
allows us to express the electron number density as
\begin{eqnarray}
  n_e 
  = 
  \mathfrak{g}_e \, \lambda_e^{-\nu} \, \frac{1}{\Gamma(\nu/2)} \,
  \int_0^\infty \frac{dx}{x} \, 
  \frac{x^{\nu/2}}{ e^{-\beta_e \mu_e} \, e^x + 1 } \,.
\label{nedef}
\end{eqnarray}

When the quantity $-\beta_e \mu_e $ becomes very large (and positive),
Fermi-Dirac statistics pass to the Maxwell-Boltzmann limit, with the
denominator above becoming a simple exponential.  In this limit, the
\hbox{$x$-integration} becomes the standard representation of $
\Gamma(\nu/2) $, and we see that the number density in this small
fugacity limit is given by the Maxwell-Boltzmann form (\ref{MBn}), 
as it must be. 
Expanding Eq.~(\ref{nedef}) to second order in the fugacity 
$z_e = e^{\beta_e \mu_e}$
and using 
$\mathfrak{g}_e=2$ gives, for the the physical case of three dimensions,
\begin{eqnarray}
  \nu = 3 \,: \hskip0.3cm 
  n_e 
  \simeq
  \frac{2}{\lambda_e^3} \, e^{\beta_e \mu_e} \left[
  1 - \frac{e^{\beta_e \mu_e}}{2^{3/2}} \right] \,.
\label{napprox}
\end{eqnarray}
Note that the first correction, which decreases the number 
density, is simply the fugacity divided by a numerical factor 
of order unity. By small fugacity, we therefore mean that 
$e^{\beta_e \mu_e} \ll 2^{3/2} \simeq 2.8$. 

As we shall see in the following section, in a space of $\nu$ dimensions, 
the energy of two charges $e$ a distance $r$ apart is proportional to
$ e^2 / r^{\nu-2}$. Since the units of a number density $n$ are
$(\text{length})^{-\nu}$, we conclude that $e^2 \, n$ has the 
units of energy over length squared, independently of the spatial
dimension $\nu$. In particular, 
\begin{eqnarray}
  \omega_b^2 = \frac{e_b^2 \, n_b}{m_b} 
\end{eqnarray}
is the squared plasma frequency for species $b$ with the fixed
dimension of an inverse-time-squared, regardless of the spatial
dimensionality $\nu$.  The situation for the squared Debye wave number
is essentially the same, except that, as noted in
Appendix~\ref{sec:difun}, in general, this quantity is
defined in terms of the fluctuations in the number density, and so
\begin{eqnarray}
  \kappa_b^2 
  = 
  \beta_b \, e_b^2 \, 
  \frac{\partial n_b}{\partial (\beta_b\mu_b)} \,.
\label{kappabdegen}
\end{eqnarray}
For Maxwell-Boltzmann statistics, the derivative that appears here just
reproduces the particle density in accord with the fact that
classical particles are described by Poisson statistics. However,
for Fermi-Dirac statistics, one must use 
\begin{eqnarray}
  \frac{\partial n_b}{\partial (\beta_b\mu_b)} 
  &=& 
  \frac{\mathfrak{g}_b \lambda_b^{-\nu} }{\Gamma(\nu/2)} 
  \int_0^\infty \frac{dx}{x} \,
  \frac{x^{\nu/2}  e^{-\beta_b \mu_b} \, e^x }
  {\left[ e^{-\beta_b \mu_b} \, e^x + 1 \right]^2} \leq n_b .
\nonumber\\
&&
\end{eqnarray}
The inequality that appears here implies that
\begin{eqnarray}
  \kappa_b^2 \leq \beta_b \, e_b^2 \, n_b = \kappa_{b\,0}^2\,.
\end{eqnarray}

For the dilute case in three dimensions, including the first 
correction in the fugacity, the electron Debye wave number is
given by 
\begin{eqnarray}
  \nu = 3 \,: &&
\nonumber\\
  \kappa_e^2 &\simeq& \beta_e \, e^2 \frac{2}{\lambda_e^3} \,
  e^{\beta_e \mu_e} \left[1 - \frac{2}{2^{3/2}}\, 
  e^{\beta_e \mu_e} \right] 
\nonumber\\
  &\simeq& 
  \beta_e \, e^2 \, n_e \,  \left[1 - \frac{1}{2^{3/2}}\, 
  e^{\beta_e \mu_e} \right] \,.
\label{kappae}
\end{eqnarray}
In the first-order fugacity correction that appears here we
can use the lowest-order result
\begin{eqnarray}
  n_e \simeq \frac{2}{\lambda_e^3} \, e^{\beta_e\mu_e}
\label{fugacity}
\end{eqnarray}
to compute the fugacity and thus write
\begin{eqnarray}
  \nu = 3 \,: &&
\nonumber\\
  \kappa_e^2 
  &\simeq&  
  \beta_e \, e^2 \, n_e \,  \left[1 - \frac{1}{2^{3/2}}\, 
  \frac{\lambda_e^3 \, n_e}{2}  \right] \,.
\label{kappaee}
\end{eqnarray}

\section{Method}

Since the method of dimensional continuation that we shall use
is novel and perhaps subtle, we present here a pedagogical account
of its basis.

\subsection{Disparate Length Scales; Expansion Parameter}

The electron-ion energy exchange brought about by their collisions in
a plasma involves a Coulomb interaction that is Debye screened at large
distances and, as we shall see in the course of our work, cut off at
short distances by quantum effects. As we shall sketch below in
Subsection~\ref{impact} for arbitrary spatial dimensions $\nu$, the
familiar elementary description of this energy transfer for $\nu = 3$
dimensions involves the impact parameter integral
\begin{eqnarray}
  \int_{b_\text{min}}^{b_\text{max}} \, \frac{db}{b} 
  = 
  \ln\!\left\{ \frac{b_\text{max}}{b_\text{min}} \right\} \,.
\label{impacted}
\end{eqnarray}
Here $b_\text{min}$ is the minimum distance of closest approach that,
in the quantum limit that is relevant here, is set by the scale of 
the electron thermal wavelength $\lambda_e$. That is, $b_\text{min}$ is
some numerical constant times $\lambda_e$. The upper limit on the
impact parameter integral is set by the electron Debye length
$\kappa_e^{-1}$, with $b_\text{max}$ some numerical multiple of 
$\kappa_e^{-1}$.  Thus
\begin{eqnarray}
  \frac{b_\text{min}}{b_\text{max}} 
  \sim 
  \lambda_e \, \kappa_e \,.
\end{eqnarray} 
The purpose of the dimensional continuation method is to precisely
determine the numerical constants that appear here.

Our method applies when the ratio $b_\text{max} / b_\text{min}$ is
large: In this case, the dimensionless parameter $\lambda_e \kappa_e$
is small, and we shall use it as our expansion parameter. As we noted
in the previous section, the Debye wave-number $\kappa_e$ always has
the dimensions of an inverse length, even at arbitrary spatial
dimension $\nu$. Hence, $ \lambda_e \kappa_e$ is a convenient
parameter to employ in our dimensional continuation scheme because it
remains dimensionless as the number of spatial dimensions $\nu$ is
varied.  Moreover, as we shall see, it is the combination 
$\lambda_e\kappa_e$ that directly arises as our computations progress.

The electron plasma coupling strength is characterized by the ratio of
the Coulomb electrostatic energy of two electrons a Debye distance apart
divided by the temperature.  In the physical space of three dimensions,
$\nu = 3 $, this is the dimensionless parameter
\begin{eqnarray}
 \nu = 3 \,: \qquad\qquad
     g_e = \beta_e \,  \frac{e^2  \kappa_e}{4\pi} \,.
\end{eqnarray}
The perturbative expansion of plasma thermodynamic parameters involve 
a series of ascending {\em integer} powers (up to additional logarithmic
corrections) of the coupling constant $g_e$.  Except for different
conventions that can alter a trivial 
overall factor, the electron quantum Coulomb 
parameter is defined as the Coulomb energy for two electrons a thermal 
wave length apart divided by the temperature. For three dimensions, 
this reads
\begin{eqnarray}
  \nu = 3 \,: \qquad\qquad
  \eta_e = \beta_e \, \frac{ e^2}{4\pi \, \lambda_e} \,.
\label{etae}
\end{eqnarray}
Hence, in three dimensions, 
\begin{eqnarray}
  \nu = 3 \,: \qquad\qquad
  \lambda_e \kappa_e = \frac{g_e}{\eta_e} \,.
\end{eqnarray}
Thus our expansion parameter $\lambda_e\kappa_e$ is essentially the
plasma coupling parameter $g_e$, albeit divided by the quantum 
parameter $\eta_e$. Accordingly, one could equivalently work in terms
of the coupling $g_e$ as we have done in the past~\cite{bps}, but 
here it is more convenient to use $\lambda_e\kappa_e$, and so this
we shall do. 

Our work applies to fully ionized plasmas where the temperature
is large and thus the parameter $\eta_e$ is small.  The condition
that $\lambda_e \kappa_e$ be small requires that the plasma
coupling $g_e$ be even smaller than $\eta_e$.  To put this in 
perspective, we recall that even if Fermi-Dirac statistics are
required, the Debye wave-number is smaller than that given by
the Maxwell-Boltzmann form with the same temperature and density.
Hence we have
\begin{eqnarray}
  \lambda_e^2 \, \kappa_e^2 
  &\leq& 
  \lambda_e^2 \, \beta_e \, e^2 \, n_e \,,
\end{eqnarray}
where the electron number density on the right-hand side of this equation 
is given by Maxwell-Boltzmann statistics. 
Using now the number density (\ref{MBn}) in the Maxwell-Boltzmann limit
and the definition (\ref{deflambd}) of the thermal wavelength, we find 
that
\begin{eqnarray}
  \lambda_e^2 \, \beta_e \, e^2 \, n_e
  &=&  
  8 \, \pi^{1/2} \,  e^{\beta_e\mu_e} \,
\sqrt{\frac{\epsilon_\smH}{T_e}}
\nonumber\\
&=& 
  8 \, \pi^{1/2} \,  e^{\beta_e\mu_e} \,
  \sqrt{\frac{13.6 \, {\rm eV} }{T_e} }  \,.
\end{eqnarray}
Thus, even for somewhat large electron fugacities $e^{\beta_e\mu_e}$, 
the expansion parameter $\lambda_e \kappa_e$ will be small provided
the temperature is reasonably large.

To explain further the utility of $\lambda_e \kappa_e$ as the 
appropriate expansion parameter, we examine the situation when 
the spatial dimension $\nu$ departs from its physical value
$\nu=3$. In this case, as we shall soon see in the next
subsection, the Coulomb potential a distance $r$ away from a
point charge has the dependence $r^{-(\nu-2)}$.  Thus the plasma 
coupling and quantum Coulomb parameters have the form
\begin{eqnarray}
  g_e \sim \beta_e \, e^2 \, \kappa_e^{\nu -2} \,,\qquad\qquad
  \eta_e 
  \sim 
  \beta_e \, \frac{ e^2}{\lambda_e^{\nu -2} } \,,
\end{eqnarray}
and so $ g_e/\eta_e \sim (\lambda_e \kappa_e)^{\nu-2} $ \, or
\begin{eqnarray}
  \lambda_e\kappa_e 
  \sim 
  \left( \frac{g_e}{\eta_e} \right)^{1/(\nu-2)} \,.
\end{eqnarray}
This emphasizes that although the form of the coupling 
$\lambda_e\kappa_e$ that we employ here does not change as the
spatial dimension is altered, its form in terms of $g_e/\eta_e$ 
does depend upon this dimensionality.

\subsection{Idea of Dimensional Continuation}

We have already seen explicitly how a geometrical quantity,
namely the number density, can be computed in a space of
arbitrary dimensionality $\nu$. In fact, all fundamental theories
can be formulated in a world that has space of arbitrary
dimensionality. Modern quantum field theory, the mother of all
physical theory, is generally formulated for spaces of arbitrary
dimensionality in order to regulate it. [See, for example,
Ref.~\cite{lsb}.] The well known BBGKY hierarchy of coupled
equations that depicts general kinetics can obviously be written
in a space of arbitrary dimensionality $\nu$. 
For $ \nu > 3 $, the Coulomb force acts
as a short-range force; for $ \nu < 3 $, it acts a long-range
force. Although the complete BBGKY set of coupled equations for
Coulomb forces must remain valid for arbitrary spatial
dimensionality $ \nu $, it cannot be approximated by the
Boltzmann or Lenard-Balescu equations for general $ \nu $ values.
To leading order in the density, the Boltzmann equation describes
the short-distance, hard scattering correctly while the
Lenard-Balescu equation correctly describes the long-distance,
collective interactions.  Hence, to this leading order, the BBGKY
hierarchy of equations reduces to the Boltzmann equation for $
\nu > 3 $, but for $ \nu < 3 $, the BBGKY hierarchy reduces to
the Lenard-Balescu equation.  We shall see how this works out in
detail as our work progresses.

Here we introduce the idea of dimensional continuation by examining
the case of electrostatics. The Poisson equation for a point charge 
$Ze$ in $\nu$ dimensions reads
\begin{eqnarray}
  -\nabla^2 \, \phi^{(\nu)}({\bf r}) 
  = 
  Z e \, \delta^{(\nu)}({\bf r}) \,.
\end{eqnarray}
Its solution may be expressed as a Fourier integral:
\begin{eqnarray}
  \phi^{(\nu)}({\bf r}) 
  = 
  \int \frac{d^\nu k}{(2\pi)^\nu} \,
  \frac{Z e}{{\bf k}^2} \, e^{ i {\bf k} \cdot {\bf r} } \ .
\end{eqnarray}
As it stands, this integral is defined for all positive integer
dimensions $\nu$. As is well known from the theory of complex
functions, an analytic function is defined from its values on the
positive real integers, provided that the function does not diverge
rapidly at infinity in the complex plane~\cite{C}. But for our
equations, we can obtain this extension by explicit calculations. For
the case at hand, we first write
\begin{eqnarray}
  \frac{1}{{\bf k}^2} \equiv \frac{1}{k^2} 
                      = \int_0^\infty ds \, e^{- k^2 \, s} \,, 
\end{eqnarray}
and interchange integrals to encounter
\begin{eqnarray}
 && \int \frac{d^\nu k}{(2\pi)^\nu} \,
  e^{- k^2 s} \, e^{ i {\bf k} \cdot {\bf r} }
\nonumber\\ 
  &=&
  \int \frac{d^\nu k}{(2\pi)^\nu} \,
  \exp\!\left\{ - \left[{\bf k} - i \frac{{\bf r}}{2s } \right]^2 \, s\right\}
  \exp\!\left\{ - \frac{r^2}{4s} \right\}
  \nonumber\\
  &=& 
  \left( \frac{1}{4\pi s} \right)^{\nu/2}
     \exp\!\left\{ - \frac{r^2}{4s} \right\} \,.
\end{eqnarray}
Here we have completed the square and used the variable in the square
brackets as the new integration variable to obtain a product of $\nu$
ordinary Gaussian integrals whose evaluation produces the final result.
The change of variables from $s$ to $ t = r^2 / 4s $ now gives
\begin{eqnarray}
  \phi^{(\nu)}({\bf r}) 
  &=& 
  \frac{Z e}{r^{\nu -2} }\,\left(\frac{1}{\pi}\right)^{\nu/2}\,\frac{1}{4}\,
  \int_0^\infty \frac{dt}{t} \, t^{(\nu - 2)/2} \, e^{-t}
\nonumber\\
&=& 
  \frac{Z e}{r^{\nu -2} }\,\left(\frac{1}{\pi}\right)^{\nu/2}\,\frac{1}{4} \, 
  \Gamma\left( \frac{\nu-2}{2} \right) \,,
\label{nuphi}
\end{eqnarray}
since the $t$ integral is a standard representation of the gamma function.
This result now defines an electrostatic potential for any value of
$\nu$ in the entire complex plane.

As a mathematical application of this result, we note that it
gives the electric field 
\begin{eqnarray}
 {\bf E}({\bf r}) 
  = 
 \frac{Z e\,   \hat{\bf r} }{r^{\nu - 1} } \,
\left(\frac{1}{\pi}\right)^{\nu/2}\,\frac{1}{2}\, 
 \Gamma\left( \frac{\nu}{2} \right) \,.
\label{E}
\end{eqnarray}
Hence Gauss' law applied to a sphere of radius $r$,
\begin{eqnarray}
  S(r) \, \hat{\bf r} \cdot {\bf E}({\bf r}) = Z e \,,
\end{eqnarray}
informs us that this sphere has an area $S = \Omega_{\nu-1}\,r^{\nu
-1}$ where
\begin{eqnarray}
  \Omega_{\nu-1} 
  =  
  \frac{2 \,\pi^{\nu/2}}{\Gamma\left(\nu/2\right)} 
\end{eqnarray}
is the area of a $(\nu-1)$-dimensional unit hypersphere embedded in
the $\nu$-dimensional space.

\begin{figure}
\includegraphics[scale=0.40]{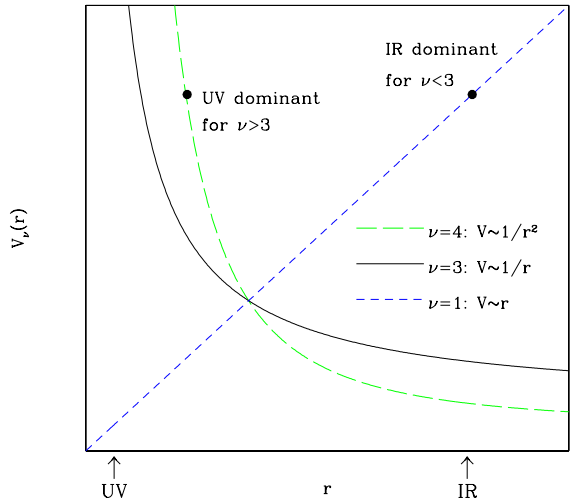}
\vskip-0.8cm 
\caption{
  Short-distance or ultraviolet (UV) physics dominates in dimensions
  $\nu>3$. Long-distance or infrared (IR) physics dominates when
  $\nu<3$. UV and IR physics are equally important in $\nu=3$.
}
\label{fig:coulomb}
\end{figure}

There are physical implications that follow from (\ref{nuphi}) of the
electrostatic potential of a point charge in $\nu$ dimensions. These
are brought out in Fig.~\ref{fig:coulomb}.
As the figure shows, the Coulomb potential of a point charge becomes
more singular at the origin as the spatial dimension $\nu$ increases:
the physics at short distances is increasingly emphasized as the
spatial dimension $\nu$ increases. Since short distances correspond to
high wave numbers, this is equivalent to stating that large $\nu$ 
emphasizes ultraviolet physical processes.  Conversely, as the spatial
dimension $\nu$ decreases, the potential falls off less rapidly at
large distances: the physics at large distances becomes ever more
important as the spatial dimension $\nu$ decreases. Since long distances
correspond to low wave numbers, this is equivalent to stating that
small $\nu$ emphasizes infrared physical processes.  As we shall see,
the electron-ion energy exchange can be computed with the Boltzmann
equation for $\nu >3$ since it correctly accounts for hard
scattering. The result, however, has a simple pole that diverges as
$\nu \to 3$ from above. Conversely, the electron-ion energy exchange
can be computed from the Lenard-Balescu equation for $\nu < 3$ since
it correctly accounts for the long-range screening brought about by
the collective, dielectric effects in the plasma. The result, however,
has a simple pole and diverges as $\nu \to 3$ from below. These general
features are brought out in the simple computation of the next
subsection.

\subsection{Energy Loss Structure in $\nu$ Spatial Dimensions}
\label{impact}

To illustrate the remarks that we have been making, we consider
the lowest-order energy loss of an electron passing by a fixed 
point of charge $Ze$.  In zeroth order, an electron with impact
parameter ${\bf b}$ and perpendicular velocity ${\bf v}$, so that
${\bf b} \cdot {\bf v} = 0$, simply follows the straight line 
${\bf b} + {\bf v} \, t $ as a function of the time $t$. In
first approximation, with the electric field ${\bf E}$ given by 
Eq.~(\ref{E}), the electron acquires a momentum transfer
\begin{equation}
\Delta {\bf p} = -e \int_{-\infty}^{+\infty} dt \,
                     {\bf E}({\bf b} + {\bf v} t )
\end{equation}
in passing by the fixed point charge $Ze$. This gives an energy change
\begin{equation}
\Delta E = \frac{{\Delta{\bf p}}^2}{2m_e} \,.
\end{equation}
It is a straightforward matter to check that this gives, up to 
numerical factors of no importance,
\begin{equation}
\Delta E \sim \frac{1}{m_e} \, \left( \frac{Z e^2}{v} \right)^2
                   \left( \frac{1}{b^{\nu - 2} } \right)^2 \,.
\end{equation}
In $\nu$ spatial dimensions, an element of cross section is given by
$d\sigma = \Omega_{\nu -2} \, b^{\nu -2} \, db $. Hence, the 
weighted energy loss has the form
\begin{equation}
\int d\sigma \, \Delta E \sim   
\int_{b_\text{min}}^{b_\text{max}} \, \frac{db}{b^{\nu - 2} } \,.
\end{equation}

This example explicitly demonstrates that large $\nu$ is dominated 
by short-distance physics and small $\nu$ is dominated by 
long-distance physics.  Moreover, it shows explicitly that 
$\nu = 3$ is the dividing line between these two regions. To bring
this out,
all we need do is to note that for $\nu > 3$ the impact parameter 
integral is not sensitive to the large distance cut off, and we may
take the limit $ {b_\text{max}} \to \infty$ to obtain
\begin{equation}
\nu > 3 \,: \qquad\qquad  
I^\smGT(\nu) =
\int_{b_\text{min}}^\infty \frac{db}{b^{\nu-2}} = 
    \frac{b_\text{min}^{3-\nu}}{\nu -3} \,.
\end{equation}
Conversely, for $\nu < 3$, we may set $ {b_\text{min}} = 0 $, with
\begin{equation}
\nu <3 \,: \qquad\qquad
I^\smLT(\nu) =
\int_0^{b_\text{max}} \frac{db}{b^{\nu-2}} = 
    \frac{b_{\text{max}}^{3-\nu}}{3-\nu} \,.
\end{equation}
The results displayed are the dominant forms in the two different
regions of spatial dimensionality $\nu$.

\subsection{Implementation of Dimensional Continuation}

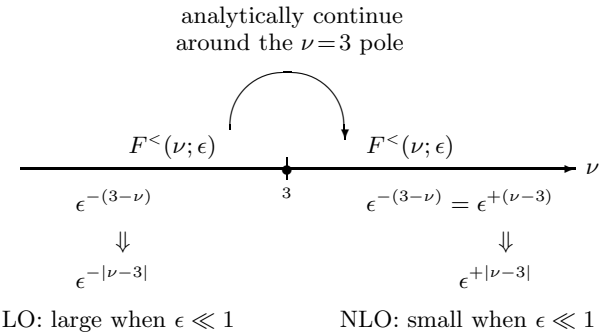
\begin{figure}[t]
\begin{picture}(120,120)(-50,0)

\put(-101,60){\vector(1,0){210}}
\put(113,58){$\nu$}

\put(0,56){\line(0,1){8}}
\put(-2.3,57.3){$\bullet$}
\put(-2.0,49){${\scriptstyle 3}$}

\put(-60,66){$F^\smLT(\nu;\epsilon)$}
\put(-80,44){$\epsilon^{-(3-\nu)}$}
\put(-65,30){$\Downarrow$}
\put(-80,15){$\epsilon^{-\vert \nu-3 \vert}$}
\put(-108,0){LO: large when $\epsilon \ll 1$}

\put(30,66){$F^\smLT(\nu;\epsilon)$}
\put(30,44){$\epsilon^{-(3-\nu)}=\epsilon^{+(\nu-3)}$}
\put(80,30){$\Downarrow$}
\put(65,15){$\epsilon^{+\vert \nu-3 \vert}$}
\put(20,0){NLO: small when $\epsilon   \ll 1$}

\put(0,75){\oval(43,43)[t]}
\put(22,72){\vector(0,-1){1}}
\put(-40,115){analytically continue}
\put(-42,105){around the $\nu\!=\!3$ pole}

\end{picture}
\caption{
The analytic continuation of $F^\smLT(\nu;\epsilon)$ from
$\nu<3$ to the region $\nu>3$: the same expression can be used
for $F^\smLT(\nu;\epsilon)$ throughout the complex plane 
since the pole at $\nu=3$ can easily be avoided. Note that the
quantity $F^\smLT(\nu;\epsilon) \sim \epsilon^{(\nu-3)}$ 
is leading order in $\epsilon$ for $\nu < 3$. However, upon analytically 
continuing to $\nu>3$ we find that 
$F^\smLT(\nu;\epsilon) \sim \epsilon^{\vert \nu - 3 \vert} $
which is next-to-leading order in $\epsilon$  
relative to $F^\smGT(\nu;\epsilon) \sim \epsilon^{-\vert \nu - 3 \vert}$.
}
\label{figure}
\end{figure}

The situation that we have just described leads to a well defined
result because it is akin to the following example.  Suppose that we
have a theory that is well defined in the neighborhood of the physical
dimension $\nu = 3$, and that the theory contains a small parameter
$\epsilon$.  Moreover, suppose that we need to evaluate a function
$F$ that depends upon this small parameter $\epsilon$ in the
following fashion.  For $\nu >3$ the leading behavior of $F$ goes like
$\epsilon^{3- \nu}$, and the function $F$ has a simple pole in $\nu$ as
$\nu \to 3$ from above. Conversely, for $\nu < 3$, the leading
behavior of the function $F$ goes like $\epsilon^{\nu-3}$ and the
function $F$ has a simple pole in $\nu$ as $\nu \to 3$ from below.
That is, we have the leading terms
\begin{equation}
\nu > 3 \,:  \qquad\qquad
F^\smGT(\nu;\epsilon) 
  = 
  A^\smGT(\nu) 
\,\epsilon^{3 - \nu} \,,
\end{equation}
and
\begin{equation}
\nu < 3 \,: \qquad\qquad
 F^\smLT (\nu;\epsilon) 
  = 
  A^\smLT(\nu) 
\,\epsilon^{\nu-3} \,.
\end{equation}
Since the two contributions each have poles in $\nu$, 
\begin{eqnarray}
  A^\smGT(\nu) 
  = 
  \frac{R^\smGT}{\nu-3} + r^\smGT + {\cal O}(\nu -3) \,,
\end{eqnarray}
and
\begin{eqnarray}
  A^\smLT(\nu) 
  = 
  \frac{R^\smLT}{3-\nu} + r^\smLT + {\cal O}(3-\nu) \,.
\end{eqnarray}
The function $F^\smLT (\nu;\epsilon) $ is of leading order in the
expansion parameter $\epsilon$ for $\nu < 3$. Since it is an analytic
function of $\nu$, it may be continued throughout the complex
$\nu$-plane. When it is analytically continued to $\nu > 3$ it becomes
of subleading order.  This behavior is depicted in Fig.~\ref{figure}.
Exactly the converse situation applies to the function 
$F^\smGT (\nu;\epsilon)$. 

Therefore, in the neighborhood of $\nu = 3$, 
\begin{eqnarray}
F(\nu;\epsilon) &=& F^\smGT (\nu;\epsilon) 
                          +  F^\smLT (\nu;\epsilon) 
\nonumber\\[5pt]
&=&  A^\smGT(\nu) \,\epsilon^{3-\nu}  +
        A^\smLT(\nu) \,\epsilon^{\nu-3}
\end{eqnarray}
is accurate to leading and sub-leading order in $\epsilon$.  For $\nu >
3$, the term with the coefficient $A^\smGT(\nu)$ is dominant while
that with the coefficient $A^\smLT(\nu)$ is sub-dominant. For $\nu <
3$ the roles of the the dominant and sub-dominant terms are
interchanged. It should be emphasized that the addition of the two
terms contains no double counting since in each region one term, and
one term only, dominates.  Since the theory is well defined at the
physical dimension $\nu = 3$, the poles must cancel, which requires
that
\begin{eqnarray}
  R^\smGT = R^\smLT \,.
\end{eqnarray}
Using
\begin{eqnarray}
  \epsilon^{\pm (\nu - 3)} = e^{\pm (\nu -3) \, \ln\epsilon} \,,
\end{eqnarray}
we now have, in the neighborhood of $\nu=3$,
\begin{eqnarray}
  F(\nu;\epsilon) 
  &=& 
  \frac{R^\smGT}{\nu -3} \, \left[ 
  e^{+ (\nu -3) \, \ln\epsilon} - e^{- (\nu -3) \, \ln\epsilon} \right]
  + r^\smGT + r^\smLT
\nonumber\\[5pt]
  &=& 
  2 R^\smGT \, \ln \epsilon + ( r^\smGT + r^\smLT )\,.
\label{ep}
\end{eqnarray}
We must emphasize that this method of dimensional continuation
provides not only the coefficient $2 R$ out in front of $\ln\epsilon$
(which is often not too difficult to compute), but the constant
$r^\smGT + r^\smLT$ in addition to this logarithm (which is often
difficult to compute).

A relevant example is provided by the simple model of the 
energy loss presented in the previous subsection. According
to our general method, in the neighborhood of $\nu=3$ we 
must have
\begin{eqnarray}
  I(\nu) &=& I^\smGT(\nu)  + I^\smLT(\nu)  
\nonumber\\[5pt]
       &=& 
    \frac{b_\text{min}^{3-\nu}}{\nu -3}
    +
   \frac{b_{\text{max}}^{3-\nu}}{3-\nu}  \,.
\end{eqnarray}
The $\nu \to 3$ limit produces
\begin{eqnarray}
I(\nu)  &=& 
    \frac{b_{\text{max}}^{3- \nu}}{\nu-3} \left[
\left( \frac{b_\text{min}}{b_\text{max}} \right)^{3 -\nu} -1 \right]
\nonumber\\[5pt]
 &\to& - \ln\!\left\{\frac{b_\text{min}}{b_\text{max}}\right\} \,.
\label{pe}
\end{eqnarray}
This is precisely the value (\ref{impacted}) of the familiar impact 
integral evaluated directly in three dimensions

Another instructive example of this method is provided by an
examination of the modified Bessel function $K_\nu(z)$ for small
$\nu$ and small $z$. This is discussed in Ref's.~\cite{lfirst}
and \cite{bps}, and in more detail in Ref.~\cite{bpsex}.  These
works should be consulted if the explanation already given is not
convincing.

Although we have sketched the basic idea of our dimensional
continuation scheme leading to the result (\ref{ep}), in this paper we
shall apply it in a slightly different form, a form similar to the
more physical example that led to the result (\ref{pe}). As we have
explained, the electrostatic potential in $\nu$ spatial dimensions has
the functional form $ e / r^{\nu-2}$ so that the energy between two
point charges a distance $r$ apart is proportional to $ e^2 /
r^{\nu-2}$.  As we shall see explicitly in our work below, the
electron-ion energy exchange rate contains an over all
dimension-bearing factor $\beta_e \, e^2 $. This factor has the
dimensions of length to the power $\nu - 2$.  To define a quantity
whose physical dimension does not vary as the spatial dimension
varies, the factor $\beta_e \, e^2$ must be accompanied by a factor of
length raised to the power $3-\nu $, which gives a result that has a
constant factor of 1/length in all spatial dimensions $\nu$. For the
$\nu > 3$ contribution, a scattering term with a length cutoff given
by the electron thermal wavelength $\lambda_e$, the needed dimensional
factor is given by $\lambda_e^{3-\nu}$ as we shall explicitly find
below.  For the $\nu < 3$ contribution, a long-distance Debye screened
interaction term, the needed dimensional factor is given by
$(1/\kappa_e)^{3-\nu}$ as we shall also explicitly see below. Thus, in
all spatial dimensions near $\nu=3$, the rate has the structure
\begin{eqnarray}
  G(\nu) 
 &=& 
 \beta_e  e^2\,\left[ \lambda_e^{3 - \nu} \, B^\smGT(\nu) +
\left(\frac{1}{\kappa_e}\right)^{3-\nu} \, B^\smLT(\nu) \right]
 \,,
\nonumber\\&&
\end{eqnarray}
and for $\nu $ near $\nu=3$,
\begin{eqnarray}
  B^\smGT(\nu) = \frac{R}{\nu -3} + b_\smGT 
\end{eqnarray}
and 
\begin{eqnarray}
  B^\smLT(\nu) = \frac{R}{3 - \nu} + b_\smLT \,.
\end{eqnarray}
Writing $B=b_\smGT + b_\smLT$, we find that for $\nu$ near $\nu = 3$,
\begin{eqnarray}
  G(\nu) 
  &=& 
  \beta_e \, e^2 \, \lambda_e^{3-\nu} 
  \left[\frac{R}{\nu-3}  
  \Big\{ 1 - \left(\lambda_e \, \kappa_e \right)^{\nu-3} \Big\} + B
  \right]
\nonumber\\[5pt]
  &\to& 
  \beta_e \, e^2 \,\Big[-R \ln\!\left\{\lambda_e \, \kappa_e \right\} + 
  B\, \Big] \,,
\end{eqnarray}
in which the final line gives the $\nu = 3$ limit.

An objection could be raised that we have not shown explicitly that
larger sub-leading terms are not present. We have extracted terms that
have the generic behavior $\epsilon^{3-\nu}$ for $\nu > 3$ and
$\epsilon^{\nu-3}$ for $\nu < 3$. One might ask if there are
additional terms with a power law dependence between
$\epsilon^{3-\nu}$ and $\epsilon^{\nu-3}$.  However, simple
dimensional analysis shows that such terms of intermediate order
cannot appear.  The point is that the physics involves {\it only two}
different mechanisms that dominate at large and small scales. These
two different mechanisms involve different combinations of basic
physical parameters and hence give quite different dependencies on the
small parameter when the spatial dimension $\nu$ departs form $\nu =
3$.

\section{Convergent Kinetic Equations}

A number of authors~\cite{jh, f&b, g&d, oa} have proposed various
versions of plasma kinetic equations that have neither short nor long
range divergences.  This work is summarized in the book of
Liboff~\cite{rll}, which we shall outline here and then relate to our
method of dimensional regularization.

Liboff, in his Eq.~(2.75), writes the transport equation for a
homogeneous system such as we consider as
\begin{eqnarray}
  \frac{\partial f}{\partial t} = B_0 + L_0 - \bar R \,,
\end{eqnarray}
where $B_0$ is the Boltzmann collision integral, $L_0$ the
Lenard-Balescu integral, and $\bar R$ is a renormalization term that
cancels the singularities in $B_0$ and $L_0$~\cite{foot}.
We should note that starting off with admittedly infinite, and
therefore undefined quantities, as in Ref.~\cite{rll}, is at best a
heuristic procedure. This is to be contrasted with the renormalization
procedure performed in modern quantum field theories where the
starting point is a rigorously defined, finite theory because the
starting point is a {\em regularized} theory. At any rate, the
infinite renormalization term $\bar R$ is expressed formally as a
double integral over both impact parameters and Fourier wave numbers.
The integral over impact parameters $b$ is broken up into a large
impact parameter part $b > b_0$ and a small impact parameter part $b <
b_0 $, $ \bar R = \bar R(> b_0) + \bar R_0(< b_0) $. It is then shown
that $\bar R$ has a formal construction such that both $B_0 - \bar R(>
b_0)$ and $L_0 - \bar R(< b_0)$ are finite.

The transport equation is thus rendered finite.  Liboff concludes,
``So we find that the combination of collision integrals gives a
reasonable model for a convergent plasma kinetic equation.''   
Our goal, however, is not just to find a ``reasonable model,'' but to
{\em calculate} the Coulomb logarithm in a precise and rigorous
fashion.  To leading order in the plasma density, we shall not only compute
the coefficient out in front of the logarithm, but also the constants
that appear in addition to the logarithm.  See Ref.~\cite{bpsex}
for more details.

Although Gould and DeWitt~\cite{g&d} also separate the right-hand
side of the transport equation into three terms, they do so in
such a fashion that each of the terms is finite and well
defined. As shown in Appendix B of BPS~\cite{bps}, the
formulation of Gould and DeWitt correctly gives the constant term
as well as the leading Coulomb logarithm, and as far as these
terms are concerned, their work is mathematically equivalent to
our method of dimensional continuation. Both are accurate
to~${\cal O}(g^2)$ in the plasma coupling, and no better. The
trouble with their formulation is that it also produces a subset
of higher order terms, and there is no reason that these provide
a more accurate evaluation.  As is well known, the inclusion of
partial subsets of higher-order terms can sometimes give less
rather than more accurate results.

\section{Energy and Temperature Rates}

The rate of change in the electron energy density transported to all the
ions species vanishes when the two subsystems have the same temperature. 
Hence we may write
\begin{eqnarray}
  \frac{d{\cal E}_{e \smI}}{dt}
  =   
  -\, {\cal C}_{e \smI} \left( T_e - T_\smI \right) \,.
\label{dedtei}
\end{eqnarray}
Since energy flows from the electrons to the ions when the electrons
are hotter than the ions, ${\cal C}_{e \smI}$ is positive.
Since the total energy is conserved, the rate at which energy is transferred
from the ions to the electrons, $ d{\cal E}_{\smI e}/dt $, has the
same coefficient $ {\cal C}_{e \smI} $ but an overall sign change or, 
equivalently,
\begin{eqnarray} 
  \frac{d{\cal E}_{\smI e }}{dt}
  =   
  -\,{\cal C}_{e \smI} \left( T_\smI - T_e \right) \,.
\label{dedtie}
\end{eqnarray}

A change in the energy of a subsystem in the plasma produces a 
corresponding change in the temperature of that subsystem.  Thus, 
for the electrons,
\begin{eqnarray}
  \Delta {\cal E}_{e \smI} = c_e \, \Delta T_e \,, 
\end{eqnarray}
while for the ions
\begin{eqnarray}
  \Delta {\cal E}_{\smI  e}
  = 
  {\sum}_i \Delta {\cal E}_i = c_\smI \, \Delta T_\smI \,.
\end{eqnarray}
Here, since the plasma interactions do not change particle number
densities, the specific heats $c_e$ and $c_\smI$ are those at constant
volume. Since $\Delta {\cal E}_{e\smI} $ is an energy density, these
are the specific heats per unit volume. For a hot plasma that is not
strongly coupled, the case treated in this paper, these specific heats
are given by the familiar ideal gas results:
\begin{eqnarray}
  c_e = 3 n_e/2
  \hskip1cm \text{and}\hskip1cm 
  c_\smI = 3 n_\smI/2 \,,
\end{eqnarray}
where $n_\smI$ is the total ionic density, the number of all the ions
per unit volume.  Thus Eq's.~(\ref{dedtei}) and (\ref{dedtie}) are 
equivalent to
\begin{eqnarray} 
  \frac{d T_e }{dt}
  = 
  -\gamma_{e \smI}\, \left( T_e - T_\smI \right) \,,
\\[-8pt]\nonumber
\label{dTe}
\end{eqnarray}
with $\gamma_{e \smI} = {\cal C}_{e \smI}/c_e$; and
\begin{eqnarray} 
  \frac{d T_\smI}{dt} = -\gamma_{\smI e}\,\left( T_\smI - T_e \right) \,,
\end{eqnarray}
with $\gamma_{\smI e} =  {\cal C}_{e \smI}/c_\smI$.
Moreover, the rate at which the separate temperatures approach one
another is given by
\begin{eqnarray}
  \frac{d\left( T_e - T_\smI \right)}{dt} 
  = 
  -\Gamma\,\left( T_e - T_\smI \right) \,,
\end{eqnarray}
in which
\begin{eqnarray}
  \Gamma 
  = 
  {\cal C}_{e \smI} \left( \frac{1}{c_e} + \frac{1}{c_\smI} \right) \,.
\end{eqnarray}
We turn now to compute the rate coefficient ${\cal C}_{e \smI}$.

\section{Boltzmann Equation: Short-Distance Physics}

We first work in $\nu>3$ dimensions where the short-distance physics
dominates. Thus the rate of change of the electron distribution is
described by the Boltzmann equation with Fermi-Dirac statistics for the
electrons and Maxwell-Boltzmann statistics for the heavy ions. The
Boltzmann equation for the electron distribution, including the Pauli
blocking of the scattered electrons and two-body quantum scattering
effects, reads
\begin{widetext}
\begin{eqnarray}
\nonumber
  \frac{\partial f_e ({\bf p}_e)}{\partial t} 
  &=&
  {\sum}_i
  \int \frac{d^\nu p_e^\prime}{(2\pi\hbar)^\nu}\,
  \frac{d^\nu p_i^\prime}{(2\pi\hbar)^\nu}\,
  \frac{d^\nu p_i}{(2\pi\hbar)^\nu}\,\big\vert\, T\, \big\vert^2\, 
\\ && 
\nonumber
  (2\pi\hbar)^\nu\,\delta^\nu\!\Big( {\bf p}_e^\prime + {\bf p}_i^\prime - 
  {\bf p}_e - {\bf p}_i \Big)\,
  (2\pi\hbar) \, \delta\!\left( \frac{{\bf p}_e^{\prime\,2}}{2m_e} + 
  \frac{{\bf p}_i^{\prime\,2}}{2m_i} - \frac{{\bf p}_e^2}{2m_e} 
   - \frac{{\bf p}_i^2}{2m_i}   \right)
\\ &&
  \bigg\{
  f_e({\bf p}_e^\prime) f_i({\bf p}_i^\prime)\Big[1-f_e({\bf p}_e)\Big] 
  -
  f_e({\bf p}_e) f_i({\bf p}_i)\Big[1-f_e({\bf p}_e^\prime)\Big] 
  \bigg\} \ ,
\label{BEfe}
\end{eqnarray}
where $T$ is the amplitude for the two-body scattering collision $e\,i
\to e^{\,\prime}\,i^{\,\prime}$, and we have omitted the spatial
convection term on the left-hand side because we are concerned with
spatially uniform plasmas.  Since the electrons are in thermal
equilibrium with each other, there is no electron-electron
interaction contribution to this time derivative. The
electron kinetic energy density --- the electron energy per unit
volume --- has the same form as the number density (\ref{fbnorm}) save
that an additional factor of $ E_e({\bf p}) = p^2 / 2m_e $ appears in
the integrand. Hence the rate at which this energy density changes
because of the electron ion interactions is given by
\begin{eqnarray}
  \frac{\partial{\cal E}_{e \smI}}{\partial t}
  =
  2\!\int\!\frac{d^\nu p_e}{(2\pi\hbar)^\nu}\, 
  \frac{p_e^2}{2 m_e}\,
  \frac{\partial f_e ({\bf p}_e)}{\partial t} \,,
\label{dedteiA}
\end{eqnarray}
where the factor of 2 multiplying the integral accounts for the
electron spin degeneracy.
Using the crossing symmetry ${\bf p}_e
\leftrightarrow {\bf p}_e^\prime$ and ${\bf p}_i \leftrightarrow {\bf
p}_i^\prime$ of the scattering amplitude $T$ in (\ref{BEfe}), the rate
of energy exchange from the electrons to the ions (\ref{dedteiA}) can
be written as
\begin{eqnarray}
\nonumber
  \frac{\partial{\cal E}_{e \smI}^\smGT}{\partial t}
  &=&
  2 \, {\sum}_i \, \int 
  \frac{d^\nu  p_e^\prime}{(2\pi\hbar)^\nu}\,
  \frac{d^\nu  p_i^\prime}{(2\pi\hbar)^\nu}\,
  \frac{d^\nu  p_e}{(2\pi\hbar)^\nu}\, 
  \frac{d^\nu  p_i}{(2\pi\hbar)^\nu}\,\big\vert\,T\,\big\vert^2\, 
\\ &&
\nonumber
  (2\pi\hbar)^\nu\,\delta^\nu\!\Big( {\bf p}_e^\prime + {\bf p}_i^\prime - 
  {\bf p}_e - {\bf p}_i \Big)\,
  (2\pi\hbar)\, \delta\!\left( 
  \frac{{\bf p}_e^{\prime\,2} - {\bf p}_e^2}{2m_e} 
  + 
  \frac{{\bf p}_i^{\prime\,2} - {\bf p}_i^2}{2m_i} \right)
\\ &&
  \frac{{\bf p}_e^{\prime\,2} - {\bf p}_e^2}{2 m_e}\,
  f_e({\bf p}_e) f_i({\bf p}_i)\Big[1-f_e({\bf p}_e^\prime)\Big] \,,
\label{dedteiBB}
\end{eqnarray}
where the factor of two in front of the sum is the spin-degeneracy
$\mathfrak{g}_e=2$ for electrons.  We have placed a ``greater-than''
superscript on the left-hand side of the equation since we are now
computing the $\nu > 3$ contribution.  We start by performing the
${\bf p}_i^{\,\prime}$-integration in Eq.~(\ref{dedteiBB}), using the
momentum conserving delta-function to set
\begin{eqnarray}
  {\bf p}_i^\prime ={\bf p}_e + {\bf p}_i - {\bf p}_e^\prime \,.
\end{eqnarray}
Defining the momentum transfer by 
\begin{eqnarray}
  {\bf q}   \equiv
  {\bf p}_e^\prime - {\bf p}_e 
  =
  {\bf p}_i - {\bf p}_i^\prime \,,
\label{qdef}
\end{eqnarray}
and the average of the initial and final electron momenta by
\begin{eqnarray}
  \bar{\bf p}   \equiv 
  \frac{1}{2}\,\left[\,{\bf p}_e^\prime + {\bf p}_e \right] \,,
\label{pdef}
\end{eqnarray}
we can simplify Eq.~(\ref{dedteiBB}) to read 
\begin{eqnarray}
\nonumber
  \frac{\partial{\cal E}_{e \smI}^\smGT}{\partial t}
  &=&
  2 \, {\sum}_i \, \int
  \frac{d^\nu p_e^\prime}{(2\pi\hbar)^\nu}\,
  \frac{d^\nu p_e}{(2\pi\hbar)^\nu}\, 
  \frac{d^\nu p_i}{(2\pi\hbar)^\nu} \, 
  \big\vert \, T \, \big\vert^2 \, 
  (2\pi\hbar)\,\delta\!\left(\frac{1}{m_i}{\bf p}_i\!\cdot\! {\bf q} -
  \frac{{1}}{m_e}\,\bar{\bf p}\!\cdot\!{\bf q} 
   - \frac{1}{2m_i} \, {\bf q}^2 \right)\,
\\[5pt] && \hskip4cm 
  \frac{1}{m_e}\,\bar{\bf p}\!\cdot\! {\bf q}~
  f_e({\bf p}_e) f_i({\bf p}_i)\Big[1-f_e({\bf p}_e^\prime)\Big] \ .
\label{dedteiDD}
\end{eqnarray}
\end{widetext}

Since $T$ is a two-body scattering amplitude, its general form can
depend upon both the square of the momentum transfer $q^2={\bf q}\cdot
{\bf q}$ and the
total center-of-mass energy $W\!=\!p^2/2m_{ei}$, where the relative
momentum is given by \hbox{ ${\bf p} \!=\!m_{ei}({\bf v}_e \!-\! {\bf
v}_i)$}, with $m_{ei}$ being the reduced electron-ion mass.  It is the
$W$-dependence in $T\!=\!T(W,q^2)$ that renders the integrals in
Eq.~(\ref{dedteiDD}) difficult to calculate because $W$ depends
explicitly on ${\bf p}_i$. In Section~12 of Ref.~\cite{bps}, this
calculation is performed to all orders in the Coulomb scattering.
  For the work here, we shall be less general and
exploit the fact that the electron-ion mass ratio $m_e/m_i$ is very
small (so the reduced mass $m_{ei}$ is almost equal to the electron
mass $m_e$). We shall assume that the electron and ion temperatures are not
orders of magnitude apart, a mild restriction in all practical
applications, so that 
\begin{eqnarray}
  \beta_e \, m_e \ll \beta_\smI \, m_i \,.
\label{mild}
\end{eqnarray}
Under these circumstances, the thermal average electron velocity is
much larger than the ion velocity, and to a very good approximation $
|{\bf v}_e - {\bf v}_i| = |{\bf v}_e|$. Thus the quantum Coulomb
parameter $\eta_{ei} = e\, e_i / 4\pi\hbar |{\bf v}_e - {\bf v}_i|$ that
appears in the Boltzmann equation can be replaced by a Coulomb
parameter that contains only the electron velocity, $\eta_{ei} \to Z_i
e^2 /4\pi \hbar |{\bf v}_e|$, where we have written $e_i = Z_i e$.
The size of this parameter is estimated by its thermal average, which
we denote by an overline. We use the simple Maxwell-Boltzmann
distribution to estimate this average.  For this classical
distribution, the thermal average of $1 / {\bf v}_e^2 $ is precisely
$m_e/T_e$, and so
\begin{equation}
\overline{\eta_{ei}^2} 
   \simeq  Z_i^2 \, \left(\frac{e^2}{4\pi\hbar}\right)^2 \, 
     \frac{m_e} {T_e} = Z_i^2 \, 2\pi \, \eta_e^2 \,, 
\end{equation}
where in the second equality we have used the previous definition
(\ref{etae}) of the electron quantum Coulomb parameter $\eta_e$ together 
with the definition (\ref{deflambd}) of the electron thermal wave length
$\lambda_e$. Another way to write this is
\begin{equation}
\overline{\eta_{ei}^2} 
   \simeq  Z_i^2 \, \frac{2 \epsilon_\smH}{T_e} \,,  
\label{squared}
\end{equation}
where $\epsilon_\smH \simeq 13.6$ eV previously noted in 
Eq.~(\ref{bind}) is the binding energy of the hydrogen atom.  
The result (\ref{squared}) demonstrates that $\eta_{ei}$ is quite small
for the elevated temperature range that concerns us. Hence the scattering
amplitude in Eq.~(\ref{dedteiDD}) can be calculated in the Born
approximation~\cite{born},
\begin{eqnarray}
  T \simeq
  T_\smB(q^2) = \hbar\, \frac{e\,e_i}{q^2} \,,
\label{TBorn}
\end{eqnarray}
a quantity that depends only upon the square of the momentum transfer
$q^2$, and not on the center-of-mass energy~$W$. 

In the Born approximation, the initial ion momentum ${\bf p}_i$
appears only in the delta-function and phase-space density explicitly
shown in Eq.~(\ref{dedteiDD}), and not in the amplitude $T_\smB(q^2)$,
and so the integration over this momentum variable can be carried
out. If it were not for the delta-function factor, the
\hbox{${\bf p}_i$-integration} would simply entail
\begin{eqnarray}
  \int \frac{d^\nu p_i}{(2\pi\hbar)^\nu} \, f_i({\bf p}_i)
  &=&    
  \int  \frac{d^\nu p_i}{(2\pi\hbar)^\nu} \,   
  \exp\!\left\{ -\beta_\smI \left[ \frac{p_i^2}{2m_i} - \mu_i \right]\right\}
\nonumber\\[5pt]
  &=&  
  \lambda_i^{-\nu} \, e^{\beta_\smI \, \mu_i }
  = 
  n_i / \mathfrak{g}_i \,.
\label{intfi}
\end{eqnarray}
Following the convention exhibited in Eq.~(\ref{fbnorm}), 
the species index $i$ for ions implicitly includes spin degrees of
freedom, and so the integration over a single 
$f_i({\bf p}_i)$ produces $n_i/\mathfrak{g}_i$.  The delta-function in
Eq.~(\ref{dedteiDD}) removes one of the components of the
$p_i$-integration, which is equivalent to supplying an extra factor of
$\lambda_i$ and retaining a Maxwell-Boltzmann factor corresponding to
the component of the momentum ${\bf p}_i$ along the direction of ${\bf
q}$.  Hence
\begin{eqnarray} 
  &&  
  \int \frac{d^\nu p_i}{(2\pi\hbar)^\nu} f_i({\bf p}_i) \,
  (2\pi\hbar)\,\delta\!\left(
  \frac{{\bf p}_i\!\cdot\!{\bf q}}{m_i}
  - 
  \frac{\bar{\bf p}\!\cdot\!{\bf q}}{m_e} 
  - 
  \frac{q^2}{2m_i}  \right) 
\nonumber\\[5pt]
  && 
  =
  \frac{1}{q}\, \frac{n_i}{\mathfrak{g}_i}\,
  \lambda_i\, m_i \exp\!\left\{-\frac{\beta_\smI}{2 m_i\,q^2} 
  \left( \frac{m_i}{m_e}  \bar{\bf p} \cdot {\bf q} + 
  \frac{q^2}{2}    \right)^2 \right\} .
\nonumber\\[3pt]
&&
\end{eqnarray}
We shall often denote the magnitude of the momentum transfer by
$q=\vert {\bf q}\vert$, as we have done here. 
We now change the remaining two integration
variables ${\bf p}_e^\prime$ and ${\bf p}_e$ in Eq.~(\ref{dedteiDD})
to the variables $\bar{\bf p}$ and ${\bf q}$ defined in
Eq's.~(\ref{qdef}) and (\ref{pdef}), a change that has a unit
Jacobian. Since the electrons are described by the Fermi-Dirac
distribution (\ref{defFeA}), the Pauli blocking term in
Eq.~(\ref{dedteiDD}) can be written as
\begin{eqnarray}
 && 1 - f_e(\bar{\bf p} + {\bf q}/2) 
\\[5pt] \nonumber
 && =
  e^{ - \beta_e \mu_e } \, 
  \exp\!\left\{\frac{\beta_e}{2m_e} \,  \left( \bar{\bf p} + \frac{1}{2} 
  {\bf q} \right)^2 \right\} \, f_e(\bar{\bf p} + {\bf q}/2) \,.
\end{eqnarray}
Using these results, and neglecting terms involving the very small
ratios $m_e/m_i$ and $\beta_e m_e/\beta_\smI m_i$, we find that 
\begin{widetext}
\begin{eqnarray}
  \frac{\partial{\cal E}_{e \smI}^\smGT}{\partial t}
  &=&
  2 \, {\sum}_i \, \frac{n_i}{\mathfrak{g}_i} \, \int 
  \frac{d^\nu {\bar p}}{(2\pi\hbar)^\nu}\,
  \frac{d^\nu q}{(2\pi\hbar)^\nu}\, 
  \big\vert \, T_\smB(q^2) \, \big\vert^2 \, 
  \frac{ m_i \lambda_i }{m_e}  \, 
 \, e^{ - \beta_e \mu_e}  
\nonumber\\
&& \qquad\qquad
  f_e(\bar{\bf p} - {\bf q}/2) \, f_e(\bar{\bf p} + {\bf q}/2) \,
  \exp\!\left\{+\,\frac{\beta_e}{2 m_e} 
  \left[ {\bar p}_\smPerp^{\,2} + \frac{1}{4}\, q^2 \right]\right\} 
\nonumber\\
&& \qquad\qquad
  \bar{\bf p} \cdot \hat{\bf q} \,
  \exp\!\left\{ -\frac{\beta_\smI}{2 m_e} \, \frac{m_i}{m_e} \,
  \left[ \bar{\bf p} \cdot \hat{\bf q}  + 
  \frac{m_e}{2 m_i}\left(1 -\frac{\beta_e}{\beta_\smI}\right)
  q \right]^2 \right\} \,,
\label{midwork}
\end{eqnarray}
\end{widetext}
where $\hat{\bf q}={\bf q}/\vert {\bf q}\vert$, 
the variable $\bar{\bf p}_\smPerp$ in the first
exponent is the component of $\bar{\bf p}$ orthogonal to the momentum
transfer ${\bf q}$, so that $\bar{\bf p}={\bar{\bf p}}_\smPerp +
(\bar{\bf p} \cdot \hat{\bf q})\,\hat{\bf q}$ with $\bar{\bf
p}_\smPerp \cdot {\bf q} = 0$.

We can simplify the rate (\ref{midwork}) by further exploiting the
consequences of the very small ratio $ m_e/m_i$. In the order of
magnitude estimates that follow, we will use the symbol $\beta$ to
designate the inverse temperatures of either the electrons or the ions. This
is possible because the temperature disparity is not very severe. We
now see that the thermal distribution functions in the second line of
Eq.~(\ref{midwork}) restrict the size of the momenta to be of the
order
\begin{eqnarray}
  {\bar p}_\smPerp^{\,2} 
  \sim 
  \frac{m_e}{\beta}
  \hskip1cm\text{and}\hskip1cm 
  q^2  \sim \frac{m_e}{\beta} \,.
\end{eqnarray}
For the longitudinal component of the electron momentum, the form of
the exponential in the last line of Eq.~(\ref{midwork}) motivates the
change of variables to
\begin{eqnarray}
  {\bar p}_\smParallel^{\,\prime}
  \equiv
  \bar{\bf p} \cdot \hat{\bf q} 
  +
  \frac{m_e}{2 m_i} \,\left(1 - 
  \frac{\beta_e}{\beta_\smI } \right) q \ .
\label{plong}
\end{eqnarray}
Under this change of variables, the last line in Eq.~(\ref{midwork})
becomes
\begin{eqnarray}
 && \left[ {\bar p}_\smParallel^{\,\prime}   - 
  \frac{m_e}{2 m_i}\left(1 -\frac{\beta_e}{\beta_\smI}\right)q \right] \,
  \exp\!\left\{ -\frac{\beta_\smI}{2 m_e} \, \frac{m_i}{m_e} \,
  {\bar p}_\smParallel^{\,\prime\,2} \right\} \,,
\nonumber\\
&&
\label{damp}
\end{eqnarray}
a term whose exponent restricts the size of the longitudinal component
to be
\begin{eqnarray}
  \vert {\bar p}_\smParallel^{\,\prime}\, \vert
  \sim  
  \sqrt\frac{m_e^2}{\beta \, m_i} 
  \sim 
  \sqrt\frac{m_e}{m_i} \,\, {\bar p}_\smPerp 
  \sim 
  \sqrt\frac{m_e}{m_i} \,\, q \ .
\end{eqnarray}
This means that the second term in square brackets at the start of
expression~(\ref{damp}), the term $(m_e/m_i)q \sim \sqrt{m_e^3/ \beta
m_i^2}$\,, is a factor $\sqrt{m_e/m_i}$ smaller than the first term
${\bar p}_\smParallel^{\,\prime}$\,. However, as we shall find, the
first term integrates identically to zero, leaving the ostensibly
smaller second term as the leading order contribution. To see this, we
first note that the electron distributions $f_e(\bar{\bf p} \mp {\bf
q}/2)$ in Eq.~(\ref{midwork}) are functions of the dimensionless
variables
\begin{eqnarray}
  \beta_e \, E_e(\bar{\bf p} \mp {\bf q}/2) 
  = 
  \frac{\beta_e}{2m_e} \, (\bar{\bf p} \mp {\bf q}/2)^2 \,.
\label{expdef}
\end{eqnarray}
Here, we must express the old variable ${\bar{\bf p}}$
in terms of the new variable 
\begin{eqnarray}
  {\bar{\bf p}}^\prime 
  &\equiv& 
  {\bar{\bf p}}_\smPerp + {\bar p}_\smParallel^{\,\prime}\,
  {\hat{\bf q}} \ ,
\label{pprimedef}
\end{eqnarray}
or in terms of the vectors~(\ref{qdef}) and (\ref{pdef}), 
\begin{eqnarray}
  {\bar{\bf p}}^\prime
  &=&
  \bar{\bf p} + \frac{m_e}{2 m_i}\left(1 - \frac{\beta_e}{\beta_i} 
  \right) {\bf q} \,.
\label{pnewvec}
\end{eqnarray}
Then from Eq.~(\ref{pnewvec}), we see that replacing the old variable
$\bar{\bf p}$ in Eq.~(\ref{expdef}) by the new variable ${\bar{\bf
p}}^{\,\prime}$ incurs relative error of order
\begin{eqnarray}
  (m_e/m_i){\bar {\bf p}}^{\,\prime} \cdot{\bf q} \, (1/{\bar p}^{\,2})
  \sim 
  (m_e/m_i)^{3/2} \ ,
\end{eqnarray}
an error beyond the leading term that we retain. That is to say, we
can simply replace
\begin{widetext}
\begin{eqnarray}
  f_e(\bar{\bf p} - {\bf q}/2)  f_e(\bar{\bf p} + {\bf q}/2) 
  \to
  f_e({\bar{\bf p}}^{\prime} - {\bf q}/2)
  f_e({\bar{\bf p}}^{\prime} + {\bf q}/2) .
\end{eqnarray}
This product is explicitly even in ${\bf q}$, as are the remaining
terms in the integrand, and consequently, the odd term ${\bar
p}_\smParallel^{\,\prime}$ in the prefactor of (\ref{damp})
integrates to zero. The energy rate (\ref{midwork}) now reduces to
\begin{eqnarray}
  \frac{\partial{\cal E}_{e \smI}^\smGT}{\partial t}
  &=&
  -  2 \, {\sum}_i \, \frac{n_i}{\mathfrak{g}_i} \, \int 
  \frac{d^\nu {\bar p}^{\,\prime}}{(2\pi\hbar)^\nu}\,
  \frac{d^\nu q}{(2\pi\hbar)^\nu}\, 
  \big\vert \, T_\smB(q^2) \, \big\vert^2 \, 
  \frac{ m_i \lambda_i }{m_e}  \, 
 \, e^{ - \beta_e \mu_e}  
\nonumber\\[5pt]
&& \qquad\qquad
  f_e({\bar{\bf p}}^{\,\prime} - {\bf q}/2)\, 
  f_e({\bar{\bf p}}^{\,\prime} + {\bf q}/2) \,
  \exp\!\left\{   +  \frac{\beta_e}{2 m_e} 
  \left[ {\bar p}_\smPerp^{\,2} + \frac{1}{4}\, q^2 \right] 
  \right\} 
\nonumber\\[5pt]
&& \qquad\qquad
  \frac{m_e}{2 m_i}\left(1 -\frac{\beta_e}{\beta_\smI}\right)q\,
  \exp\!\left\{ -\frac{\beta_\smI}{2 m_e} \, \frac{m_i}{m_e} \,
  {\bar p}_\smParallel^{\,\prime\,2} \right\} \,.
\label{near}
\end{eqnarray}
The integral over the momentum ${\bar{\bf p}}^{\,\prime} =
{\bar{\bf p}}_\smPerp + {\bar p}_\smParallel^\prime\,\hat{\bf q}$
contains $\nu-1$ integrals from ${\bar{\bf p}}_\smPerp$ and one
integral from~${\bar p}_\smParallel^{\,\prime}$\,.

Now that the leading contribution has be extracted, we can make
further reductions by omitting several more terms in $m_e/m_i$\,. In
particular, we may now neglect the longitudinal part ${\bar
p}_\smParallel^{\,\prime} = {\bar{\bf p}}^{\,\prime} \cdot \hat{\bf
q}$ relative to $q=\vert{\bf q}\vert$ in the electron distribution
functions $f_e({\bar{\bf p}}^{\,\prime} \mp {\bf q})$, which then
become functions only of $\left({\bar{\bf p}}_\smPerp \mp {\bf
q}/2\right)^2$\,. In fact, since ${\bar{\bf p}}_\smPerp\cdot {\bf q} =
0$, both electron distribution functions have the same argument, 
\begin{eqnarray}
  \beta_e \, E_e({\bar{\bf p}}_\smPerp \pm {\bf q}/2) 
  =
  \frac{\beta_e}{2m_e} \, \left( {\bar p}_\smPerp^{\,2} + 
  \frac{1}{4} \, q^2 \right) \,,
\end{eqnarray}
and their product becomes a simple square: $f_e({\bar{\bf
p}}^{\,\prime} - {\bf q}/2) \, f_e({\bar{\bf p}}^{\,\prime} + {\bf
q}/2)=[f_e({\bar{\bf p}}_\smPerp + {\bf q}/2)]^2$. The longitudinal
component ${\bar p}_\smParallel^{\,\prime}$ now appears only in the
final factor of the integrand in Eq.~(\ref{near}), and we may
therefore explicitly perform the integration over this part of the
momentum,
\begin{eqnarray}
  \int_{-\infty}^\infty \frac{d {\bar p}_\smParallel^{\,\prime}}
  {2\pi\hbar} \, \exp\!\left\{ -\frac{\beta_\smI}{2 m_i} \,
  \left(\frac{m_i}{m_e} \right)^2 {\bar p}^{\,\prime\,2}_\smParallel 
  \right\} &=& \frac{m_e}{m_i\,\lambda_i} \,,
\end{eqnarray}
where the ionic thermal wave-length $\lambda_i$ is determined from
Eq.~(\ref{deflambd}). We can now express the rate (\ref{near}) as 
\begin{eqnarray}
  \frac{\partial{\cal E}_{e \smI}^\smGT}{\partial t}
  &=&
  -m_e \,\left(1 -\frac{\beta_e}{\beta_\smI} \right) \,
  {\sum}_i \, \frac{n_i}{\mathfrak{g}_i \, m_i} \, \int 
  \frac{d^{\nu-1} {\bar p}_\smPerp}{(2\pi\hbar)^{\nu-1} } \,
  \frac{d^\nu q}{(2\pi\hbar)^\nu}\, 
  \big\vert \, T_\smB(q^2) \, \big\vert^2 \, q
\nonumber\\[5pt]
&& \qquad\qquad
  e^{ - \beta_e \mu_e} \, \Big[f_e({\bar{\bf p}}_\smPerp + {\bf q}/2)\Big]^2 \,
  \exp\!\left\{   +  \frac{\beta_e}{2 m_e} 
  \left[ {\bar p}_\smPerp^{\,2} + \frac{1}{4}\, q^2 \right] 
  \right\} \,.
\label{nearer}
\end{eqnarray}
The integration over the momentum transfer ${\bf q}$ is damped at large
values of $q=\vert{\bf q}\vert$, because at such large values
\begin{eqnarray}
&& q \to \infty \,:
\nonumber\\
&&
  e^{-\beta_e \mu_e} \, 
  \Big[f_e({\bar{\bf p}}_\smPerp + {\bf q}/2)\Big]^2 \,
  \exp\!\left\{+\,\frac{\beta_e}{2 m_e} 
  \left[ {\bar p}_\smPerp^{\,2} + \frac{1}{4}\, q^2 \right] \right\}
  \to  
  e^{+ \beta_e \mu_e}\,\exp\!\left\{-\,\frac{\beta_e}{2 m_e} 
  \left[ {\bar p}_\smPerp^{\,2} + \frac{1}{4}\,q^2 \right] \right\} \,.
\label{asympq}
\end{eqnarray}
Since the limit (\ref{asympq}) constrains the integrand to small-$q$,
this further supports the use of the Born Approximation (\ref{TBorn}),
which allows us to express the rate (\ref{nearer}) as
\begin{eqnarray}
  \frac{\partial{\cal E}_{e \smI}^\smGT}{\partial t}
  &=&
  - \beta_e e^2 \, m_e\,\hbar^2 \, \omega_\smI^2 
  \Big( T_e - T_\smI \Big)  \int 
  \frac{d^{\nu-1} {\bar p}_\smPerp}{(2\pi\hbar)^{\nu-1} } \,
  \frac{d^\nu q}{(2\pi\hbar)^\nu}\, 
  \frac{1}{q^3}  
\nonumber\\[5pt]
&& \qquad\qquad
  e^{ - \beta_e \mu_e}\,\Big[f_e({\bar{\bf p}}_\smPerp + {\bf q}/2)\Big]^2\,
  \exp\!\left\{   +  \frac{\beta_e}{2 m_e} 
  \left[ {\bar p}_\smPerp^{\,2} + \frac{1}{4}\, q^2 \right] \right\} \,,
\label{nearest}
\end{eqnarray}
\end{widetext}
where $\omega_\smI^2$ is the sum over all the ionic species of the 
squared plasma frequencies, 
\begin{eqnarray}
\nonumber\\
  \omega_\smI^2 \, 
  = 
  \sum_{\rm ion \,\, species} \omega_i^2
  = 
  {\sum}_i \, \frac{e_i^2 \, n_i}{m_i} \,.
\label{ionpl}
\\\nonumber
\end{eqnarray} 

Now that the clutter has abated, we can more easily study the 
nature of the parameters that enter into
the $\nu > 3$ contribution of the energy exchange rate.
Dividing the time derivative of the electron energy 
density by the the electron specific heat $3 n_e /2 $ gives the
rate of the electron temperature change already noted in (\ref{dTe}),
namely
\begin{eqnarray}
  \frac{\partial T_e}{\partial t} 
  = 
  - \gamma_{e \smI} \, \left( T_e - T_\smI \right) \,.
\end{eqnarray}
The integral of each momentum, with the normalizing denominator 
$2\pi\hbar$, gives a pure number times a factor of
$ 1 / \lambda_e$. Since $n_e \sim \lambda_e^{-\nu} $, and each factor
of the momentum transfer $|{\bf q}|$ will produce a factor of
$ \hbar / \lambda_e$, we conclude from Eq.~(\ref{nearest}) that
\begin{eqnarray}
  \gamma_{e \smI} 
  &\sim& 
  \beta_e e^2 \, m_e \,  \hbar^2  \, \omega_\smI^2\,\,
  \lambda^{1-\nu}_e \, \left( \lambda_e / \hbar \right)^3 
\nonumber\\[5pt]
  &\sim& 
  \left( \frac{e^2}{\lambda^{\nu -2}_e} \, \frac{1}{T_e} \right)\, 
  \left( \frac{\hbar \, \omega_\smI}{T_e} \right)\, \omega_\smI \,,
\end{eqnarray}
where in the second line we have made use of 
\begin{eqnarray}
  \lambda_e^2 \sim \hbar^2 \beta_e/m_e
  ~~\text{and}~~
  \beta_e = 1/T_e \ .
\end{eqnarray}
In a \hbox{$\nu$-dimensional}
space, the energy between two electrons a distance $\lambda_e$ apart
is, up to a constant,
given by $ e^2 / \lambda_e^{\nu-2}$.  Hence the first factor in
parenthesis in the last line above is dimensionless.  Since $\hbar
\omega_\smI$ is an energy, the second factor is also dimensionless.
Thus the overall dimension of $\gamma_{e \smI}$ is that of the final
factor $\omega_\smI$, the correct dimension of an inverse time or
rate. Although a factor of $\hbar$ appears here, it is canceled by the
single factor of $\hbar$ that appears in $1/\lambda_e^{\nu-2}$ in the
$\nu \to 3$ limit, an so in this limit the rate is a classical
quantity.  However, as we shall see, the dimensional continuation
method that we use produces logarithms, and a logarithm of $\hbar$
will appear in the final result. Finally, we should note that the rate
involves the first power of the ion density, a power that does not
depend upon the spatial dimensionality $\nu$.

The rate (\ref{nearest}) for $\nu > 3$ diverges when \hbox{$\nu \to
3^+$}, a divergence that is canceled by the \hbox{$\nu \to 3^-$} limit
of the rate for $\nu < 3$ that we compute in the next section. This
latter rate involves purely classical dynamics. Thus it entails a
wave-number ${\bf k}$ that comes from the Fourier transform of a
potential which is the analog of the quantum momentum transfer ${\bf
q}$, but with ${\bf q} = \hbar \, {\bf k}$. With this replacement, the
electron distributions would become \hbox{$f_e({\bar{\bf
p}}^{\,\prime} + \hbar {\bf k}/2)$}, but since only classical
quantities appear in the forthcoming $\nu < 3$ contribution, in this
part the electron distributions must appear only as $f_e({\bar{\bf
p}}^{\,\prime})$. Thus, to separate out a part of the $\nu > 3$
Boltzmann expression for the rate that will combine in a simple
fashion with the $\nu < 3$ contribution that we shall soon examine, we
construct this part by making the replacement
\begin{eqnarray}
&&  \Big[f_e({\bar{\bf p}}_\smPerp + {\bf q}/2) \Big]^2 \,
  \exp\!\left\{ + \frac{\beta_e}{8 m_e} \, q^2 \right\} 
\nonumber\\
&&\to
  \Big[f_e({\bar{\bf p}}_\smPerp )\Big]^2 \,
  \exp\!\left\{ - \frac{\beta_e}{8 m_e} \, q^2 \right\} \,,
\\\nonumber
\end{eqnarray}
which exhibits the needed large $q^2$ damping given in the limit
(\ref{asympq}). Accordingly, we decompose the rate of energy transfer
into a potentially singular part and a regular part,
\begin{eqnarray}
  \frac{\partial{\cal E}_{e \smI}^\smGT}{\partial t} =
  \frac{\partial{\cal E}_{e \smI}^{\smGT \, \smS}}{\partial t} +
  \frac{\partial{\cal E}_{e \smI}^{\smGT \, \smR}}{\partial t} \,,
\end{eqnarray}
where
\begin{widetext}
\begin{eqnarray}
  \frac{\partial{\cal E}_{e \smI}^{\smGT \, \smS}}{\partial t}
  &=&
  -\beta_e\, e^2 \, m_e \,
  \hbar^2 \, \omega_\smI^2\, \Big( T_e - T_\smI \Big)   \int 
  \frac{d^{\nu-1} {\bar p}_\smPerp}{(2\pi\hbar)^{\nu-1} } \,
  \frac{d^\nu q}{(2\pi\hbar)^\nu}\, 
  \frac{1}{q^3} \, \exp\!\left\{ -\frac{\beta_e}{8 m_e} \, q^2  
  \right\} 
\nonumber\\[5pt]
&& \qquad\qquad\qquad\qquad
  e^{ - \beta_e \mu_e} \, \Big[f_e({\bar{\bf p}}_\smPerp )\Big]^2 \,
  \exp\!\left\{   +  \frac{\beta_e}{2 m_e} \, 
  {\bar p}_\smPerp^{\,2} 
  \right\} \,,
\label{dones}
\end{eqnarray}
and
\begin{eqnarray}
 && 
  \frac{\partial{\cal E}_{e \smI}^{\smGT \, \smR}}{\partial t}
  =
  -\beta_e\, e^2 \, m_e \,
  \hbar^2 \, \omega_\smI^2\, \Big( T_e - T_\smI \Big)  \int 
  \frac{d^2\, {\bar p}_\smPerp}{(2\pi\hbar)^{2} } \,
  \frac{d^3 q}{(2\pi\hbar)^3}\, 
  \frac{1}{q^3} \,  e^{ - \beta_e \mu_e} 
\nonumber\\[5pt]
&&   \hskip1.0cm 
  \Bigg(~ \Big[f_e({\bar{\bf p}}_\smPerp + {\bf q}/2)\Big]^2 \,
  \exp\!\left\{ + \frac{\beta_e}{2 m_e} 
  \left( {\bar p}_\smPerp^{\,2} + \frac{1}{4}\, q^2 \right)
  \right\} - \Big[f_e({\bar{\bf p}}_\smPerp )\Big]^2 \,
  \exp\!\left\{ + \frac{\beta_e}{2 m_e} 
  \left( {\bar p}_\smPerp^{\,2} - \frac{1}{4}\, q^2 \right)
  \right\} ~\Bigg) \,.
\label{doner}
\end{eqnarray}
Here in the regular part, we have taken the $\nu \to 3$ limit since
there is no impediment in so doing.

The singular part (\ref{dones}) may be simplified by performing the
$q$-integration. Passing to hyper-spherical coordinates gives
\begin{eqnarray}
  \int  \frac{d^\nu q}{(2\pi\hbar)^\nu}\,\frac{1}{q^3}\, 
  \exp\!\left\{ -\frac{\beta_e}{8 m_e} \, q^2\right\} 
  &=&
  \frac{\Omega_{\nu-1}}{(2\pi\hbar)^\nu}\,\int_0^\infty \frac{dq}{q} \, 
  q^{\nu -3} \, \exp\!\left\{ -\frac{\beta_e}{8 m_e} \, q^2  \right\}
\nonumber\\[7pt]
  &=&
  \frac{\Omega_{\nu-1}}{(2\pi\hbar)^\nu} \,\frac{1}{2} 
  \int_0^\infty \frac{dx}{x} \, 
  \left(\frac{8 m_e}{\beta_e} \, x \right)^{(\nu -3)/2 } \!\! e^{-x}
\nonumber\\[7pt]
  &=&  
  \frac{\Omega_{\nu-1}}{(2\pi\hbar)^3} \, \frac{1}{2} \, 
  \left(\frac{\pi\lambda_e^2}{4}\right)^{(3-\nu)/2} \,
  \Gamma\left(\frac{\nu -3}{2}\right) \,, 
\end{eqnarray}
where $\Omega_{\nu-1} $ is the area of a unit $(\nu\!-\!1)$-sphere
embedded in a $\nu$-dimensional space. In the second line above we
have made an obvious change to a dimensionless integration variable
$x$, and in the last line we have identified the resulting integral
with a standard representation of the gamma function.  We thus have
\begin{eqnarray}
  \frac{\partial{\cal E}_{e \smI}^{\smGT \, \smS}}{\partial t}
  &=&
  -\frac {\beta_e\,  e^2 \, m_e}{2\hbar} ~ \omega_\smI^2 \,  
  \frac{\Omega_{\nu-1}}{(2\pi)^3} \,
  \left( \frac{\pi\lambda_e^2}{4} \right)^{(3-\nu)/2}\, 
  \Gamma\left(\frac{\nu -3}{2}\right) \Big( T_e - T_\smI \Big) \,
\nonumber\\[5pt] 
&& \qquad\qquad
   \int \frac{d^{\nu-1} {\bar p}_\smPerp}{(2\pi\hbar)^{\nu-1} } \,
   \Big[f_e({\bar{\bf p}}_\smPerp )\Big]^2 \,
  \exp\!\left\{\beta_e\left[\frac{{\bar p}_\smPerp^{\,2} }{2 m_e} -
  \mu_e \right]
  \right\} \,.
\label{donees}
\end{eqnarray}
As we shall see, the $\nu<3$ contribution calculated in the next
section contains the same integral over the transverse components
${\bar{\bf p}}_\smPerp$, so it will be convenient to perform this
integral when we add these terms together.

The regular part of the energy exchange rate may also be simplified
since the integral over the electron distribution functions can be
performed when $\nu=3$. Namely, we pass to polar coordinates, with the
angular integration simply producing a factor of $2\pi$, to obtain
\begin{eqnarray}
  && 
  \int \frac{d^2 {\bar p}_\smPerp}{(2\pi\hbar)^{2} } \,
  \Big[f_e({\bar{\bf p}}_\smPerp )\Big]^2 
  \exp\!\left\{\beta_e\,\left[\frac{{\bar p}_\smPerp^{\,2}}{2 m_e}  - 
  \mu_e \right] \right\} 
\nonumber\\[5pt]
&& \qquad\qquad 
  = 
  \frac{m_e}{2\pi\beta_e \, \hbar^2} \int_0^\infty
  d\left(\frac{\beta_e \, p^2}{2m_e} \right) \, 
  \frac{ \exp\!\left\{ + \beta_e \,
  \left[\frac{p^2}{2 m_e}  - \mu_e \right] \right\}}
  {\left[  \exp\!\left\{ + \beta_e \,
  \left[\frac{p^2}{2 m_e}  - \mu_e \right] \right\} \,+  1  \right]^2 } 
\nonumber\\[5pt]
&& \qquad\qquad 
  = 
  \frac{1}{\lambda_e^2} \,
  \frac{1}{ \exp\!\left\{ - \beta_e \, \mu_e \right\}  +  1 } \,.
\label{2d}
\end{eqnarray}
Hence, making the replacement 
\begin{eqnarray}
  - \beta_e\, \mu_e \to \beta_e \left[ \frac{q^2}{8 m_e} - \mu_e \right]
\end{eqnarray}
for the first term, we have
\begin{eqnarray}
  \frac{\partial{\cal E}_{e \smI}^{\smGT \, \smR}}{\partial t}
  &=&
  -\frac{\beta_e \,e^2 m_e}{2\hbar}\, 
  \frac{\omega_\smI^2}{\pi^2}\,\frac{1 }{\lambda_e^2} \, 
  \Big( T_e - T_\smI \Big)
  \int_0^\infty \frac{dq}{q}    
\nonumber\\[3pt]
&&  \qquad
  \left\{ \frac{1}{ \exp\!\left\{ \beta_e \, \left[
  \frac{q^2}{8m_e} - \mu_e \right] \right\}  +  1 } - 
  \frac{ \exp\!\left\{- \beta_e \, \frac{q^2}{8m_e} \right\}}
  {\exp\!\left\{ - \beta_e \, \mu_e \right\}  +  1 } \right\} 
\nonumber\\[10pt]
  &=& 
  -\frac{\beta_e\, e^2 m_e}{2\hbar}\, 
  \frac{\omega_\smI^2}{\pi^2} \,\frac{1 }{\lambda_e^2}\, 
  \Big( T_e - T_\smI \Big) \,
  \frac{1}{2} \int_0^\infty dx \, \ln x
\nonumber\\[3pt]
&&  \qquad
  \left\{ \frac{ \exp\!\left\{ x - \beta_e \, \mu_e  \right\} }
  {\left[ \exp\!\left\{ x - \beta_e \, \mu_e  \right\}  +  1 \right]^2 } 
  - \frac{ \exp\!\left\{- x \right\}}
  {\exp\!\left\{ - \beta_e \, \mu_e \right\}  +  1 } \right\} \,,
\label{finite}
\end{eqnarray}
where the last line follows from a trivial change of integration 
variables and a partial integration.

\section{\label{sec:LBE} Lenard-Balescu Equation: Long-Distance Physics}

We turn now to calculate the leading order long-distance physics by
working in spatial dimensions $\nu<3$. This is done by employing the
Lenard-Balescu equation with Fermi-Dirac statistics for the electrons
and Maxwell-Boltzmann statistics for the heavy ions.  For the spatially 
homogeneous system that we work with, the Lenard-Balescu equation with
the appropriate Pauli-blocking reads
\begin{eqnarray}
\nonumber
  \frac{\partial f_e({\bf p}_e)}{\partial t} 
  &=&
  -\frac{\partial}{\partial {\bf p}_e} \cdot 
  {\sum}_i \int \frac{d^\nu p_i}{(2\pi\hbar)^\nu}\,
  \frac{d^\nu k}{(2\pi)^\nu}\, 
  {\bf k}\, 
  \bigg\vert \frac{e\, e_i}{k^2\,\epsilon(k , {\bf k}\cdot{\bf v}_i)}
  \bigg\vert^2
  \pi\,\delta\Big({\bf k}\!\cdot\!{\bf v}_e-{\bf k}\!\cdot\!{\bf v}_i\Big)
\\[5pt] && \hskip1cm 
  \bigg\{
  {\bf k} \cdot \frac{\partial f_i({\bf p}_i)}{\partial {\bf p}_i}\,
  f_e({\bf p}_e)\Big[1 - f_e({\bf p}_e)\Big]
  -
  f_i({\bf p}_i)\, {\bf k} \cdot \frac{\partial f_e({\bf p}_e)}
  {\partial {\bf p}_e}  \bigg\} \, ,
\label{LBEfe}
\end{eqnarray}
where the gradient $\partial/\partial {\bf p}_e$ acts on everything to
its right, and $\epsilon( k,\omega)$ is the classical dielectric
function for the plasma discussed in Appendix~\ref{sec:difun}. As was shown in
Appendix C of Ref.~\cite{bps}, the usual non-degenerate Lenard-Balescu
equation is a formal limit of the Boltzmann equation. The same methods
that were employed there may be used to show that Eq.~(\ref{LBEfe}) is
the corresponding long-distance equation when the electrons are
degenerate and described by Fermi-Dirac statistics.  If the electrons
and ions are in equilibrium with themselves at temperatures $T_e$ and
$T_i$, respectively, then their distribution functions $f_e$ and $f_i$
are given by Eq's.~(\ref{defFeA}) and (\ref{defFiA}), in which case the
terms in curly braces can be written as
\begin{eqnarray}
  \Big(\beta_e \,{\bf k} \cdot {\bf v}_e - \beta_i \,{\bf k} \cdot 
  {\bf v}_i \Big)\, f_i({\bf p}_i) \, \Big[f_e({\bf p}_e)\Big]^2 
  \exp\!\left\{ \beta_e\,\left[ \frac{p_e^2}{2m_e} - \mu_e \right]\right\} \, .
\label{fesimple}
\end{eqnarray}
Because the delta-function equates ${\bf k} \cdot {\bf v}_e$ with
${\bf k}\cdot {\bf v}_i$, the factor (\ref{fesimple}) and with it 
the right-hand-side of Eq.~(\ref{LBEfe}) vanishes when 
the electrons and ions are in thermal
equilibrium with a common temperature $T$, the electrons being
described by a Fermi-Dirac distribution and the ions by a
Maxwell-Boltzmann distribution. This confirms the validity of
Eq.~(\ref{LBEfe}), the Lenard-Balescu equation with Pauli-blocking for
degenerate electrons.

Using (\ref{fesimple}) with the ions at the same inverse temperature
$\beta_\smI= 1/T_\smI$, and upon integrating the total derivative
$\partial / \partial {\bf p}_e$ by parts, we can express the 
energy exchange rate as
\begin{eqnarray}
  \frac{\partial{\cal E}_{e \smI}^\smLT}{\partial t}
  &=&
   2 \, {\sum}_i \int \frac{d^\nu p_e}{(2\pi\hbar)^\nu}\, 
  \frac{d^\nu p_i}{(2\pi\hbar)^\nu}\,\frac{d^\nu  k}{(2\pi)^\nu}\, 
  {\bf k}\cdot {\bf v}_e  \,
  \bigg\vert \frac{e\, e_i}{k^2\, \epsilon(k ,{\bf k}\cdot{\bf v}_i)}
  \bigg\vert^2 
  \pi\, \delta\Big({\bf k}\!\cdot\!{\bf v}_e-{\bf k}\!\cdot\!{\bf v}_i\Big)
\nonumber\\[5pt]
  && \qquad 
  \Big(\beta_e \,{\bf k} \cdot {\bf v}_e - \beta_\smI \,{\bf k} \cdot 
  {\bf v}_i \Big)\, f_i({\bf p}_i) \Big[f_e({\bf p}_e)\Big]^2\, 
  \exp\!\left\{ \beta_e \, 
  \left[ \frac{p_e^2}{2m_e} - \mu_e \right]\right\} \,.
\label{dedteigtB}
\end{eqnarray}
\end{widetext}
We have placed a ``less-than'' superscript on the left-hand side of
(\ref{dedteigtB}) to remind ourselves that the calculation is
performed in $\nu<3$ using the Lenard-Balescu equation.  The
distribution functions constrain the velocities of the ions and the
electrons to be of the order $v_i \sim \sqrt{T_\smI/m_i}$ and $v_e
\sim \sqrt{T_e/m_e}$, respectively.  Since an ion mass is so much
greater than that of an electron, and the temperature disparity is
never excessively large for cases of interest, the ions move much
slower than the electrons,
\begin{eqnarray}
  v_i \ll v_e \,.
\label{vvei}
\end{eqnarray}
This restriction also follows from the previous condition
(\ref{mild}), the condition that \hbox{$\beta_e m_e \ll \beta_\smI
m_i$}.  To compute the rate~(\ref{dedteigtB}), we first decompose
the electron momentum into perpendicular and longitudinal components
relative to the direction specified by $\hat{\bf k}$, so that ${\bf
p}_e={\bf p}_\smPerp + p_\smParallel\, \hat{\bf k}$ with $\hat{\bf k}
\cdot {\bf p}_\smPerp =0$ and $p_\smParallel = \hat{\bf k}\cdot {\bf
p}_e = m_e\, \hat{\bf k}\cdot {\bf v}_e $. The delta function in
Eq.~(\ref{dedteigtB}) can be used to remove the parallel component
$p_\smParallel$ of the electron momenta integration.  Since $d
p_\smParallel/(2\pi\hbar) = (m_e/2\pi\hbar k) \, d ({\bf k} \cdot {\bf
v}_e)$, the use of this delta function produces a factor $
(m_e/2\pi\hbar k) $ and makes the replacement $p_\smParallel \to
m_e\,\hat{\bf k}\cdot {\bf v}_i$.  In view of the limit (\ref{vvei}),
this replacement makes $p_\smParallel$ much smaller than the magnitude
of the perpendicular components of the electron momenta
$p_\smPerp$. Hence, we can simply replace ${\bf p} \to {\bf
p}_\smPerp$ in the remainder of the integrand.  We shall also find it
convenient (the heart hath its reasons), to insert a factor of unity
in the form
\begin{eqnarray}
  1 = \int_{-\infty}^{+\infty} \!\! dv~
  \delta\!\left( v - \hat{\bf k}\cdot {\bf v}_i  \right) \,,
\label{heart}
\end{eqnarray}
which allows us to express Eq.~(\ref{dedteigtB}) as
\begin{widetext}
\begin{eqnarray}
  \frac{\partial{\cal E}_{e \smI}^\smLT}{\partial t}
  &=&
  -2\,\beta_e \frac{e^2 \, m_e}{2\pi\hbar} \,\Big(T_e - T_\smI \Big) 
  \int \frac{d^{\nu-1}\, p_\smPerp}{(2\pi\hbar)^{\nu-1}} \, 
  \Big[f_e({\bf p}_\smPerp)\Big]^2 
  \exp\!\left\{ \beta_e \, \left[ \frac{p_\smPerp^{\,2}}{2m_e} - 
  \mu_e \right]\right\} 
\nonumber\\[5pt]
&&   
  \int\frac{d^\nu k}{(2\pi)^\nu} \, k \int_{-\infty}^{+\infty} dv \,
  \frac{\pi v^2} {|k^2 \, \epsilon( k,v k)|^2}  {\sum}_i \, \beta_\smI \, 
  e_i^2 \int\!\frac{d^\nu p_i}{(2\pi\hbar)^\nu}\,f_i({\bf p}_i) \, 
  \delta\Big(\hat {\bf k} \cdot {\bf v}_i  -  v\Big) \,.
\label{dedt<}
\end{eqnarray}

This integral can be further simplified by taking advantage of the
analytic properties of the dielectric function $\epsilon(k,\omega)$,
discussed in some detail in Appendix~\ref{sec:difun}. Repeating
Eq.~(\ref{wonder}) here for convenience, we see that a considerable
portion of the integral simplifies because
\begin{eqnarray}
&& 
  \frac{\pi v}{|k^2 \, \epsilon( k , v k )|^2} \,{\sum}_i \, 
  \beta_\smI \, e_i^2 \,\int \frac{d^\nu p_i}{(2\pi\hbar)^\nu}\,  
  f_i({\bf p}_i)\,\delta\Big(\hat{\bf k}\!\cdot\!{\bf v}_i
  - v \Big) 
\nonumber\\[5pt]
&& 
  \qquad\qquad\qquad\qquad =
  - \frac{1}{2i} \, \left\{ \frac{1}{k^2 + \kappa_e^2 + F_{\!\smI}(v)} 
  -  \frac{1}{k^2 + \kappa_e^2 + F_{\!\smI}(-v)} \right\} \,. 
\label{wonderer}
\end{eqnarray}
This result is the (unemotional) reason that the factor of unity in
the form displayed in Eq.~(\ref{heart}) was inserted in the integrand. 
Here $\kappa_e$ is the electronic contribution to the Debye wave
number, including the effects of Fermi-Dirac statistics, as expressed
by Eq.~(\ref{kefe}), while $F_{\!\smI}$ is a
complex-valued function defined 
by Eq.~(\ref{Fform}). It is important to
realize that the simplification (\ref{wonderer}) only occurs when the
ion species are summed over.  The function $F_{\!\smI}(z)$ is analytic
over the upper half of the complex $z$-plane, and has the asymptotic behavior
\begin{eqnarray}
  |z| \to \infty \,: \hskip2cm 
  F_{\!\smI}(z) \to - \frac{\omega_\smI^2}{z^2} \,,
\label{zinf}
\end{eqnarray}
where $\omega_\smI$ is the total ionic plasma frequency defined above in 
Eq.~(\ref{ionpl}).  Since an explicit odd factor of $v$ appears in the 
integrand, we can write the resulting integral over $v$ in Eq's.~(\ref{dedt<})
in the form 
\begin{eqnarray}
  -\int_{-\infty}^{+\infty} \!\! dv\, \frac{v}{2i} \,
  \left\{ \frac{1}{k^2 + \kappa_e^2 + F_{\!\smI}(v)} 
  -\frac{1}{k^2 + \kappa_e^2 + F_{\!\smI}(-v)} \right\} 
  &=& 
  \lim_{V\to \infty} i\int_{-V}^{+V}\!\! dv\,
  \frac{v}{k^2 + \kappa_e^2 + F_{\!\smI}(v)} \,.
\label{odd}
\end{eqnarray}
The delta-function in Eq.~(\ref{wonderer}) removes the longitudinal
components of the ionic momenta, leaving a Maxwell-Boltzmann factor
involving the velocity $v$. Hence the left-hand-side of the integrand
in Eq.~(\ref{odd}) is damped in a Gaussian fashion for large $|v|$.
This rapid damping results from a cancellation between the terms with
$F_{\!\smI}(v)$ and $F_{\!\smI}(-v)$ that only happens when the two
velocities are exactly the negative of one another. Hence, when we
simplify the integrand by taking advantage of the odd prefactor as
was done on the right-hand side of Eq.~(\ref{odd}), we must integrate
between the exact same negative and positive limits, between $-V$ and
$+V$, and only afterward take the limit $V\to\infty$. Since
$F_{\!\smI}(z)$ is analytic in the upper-half \hbox{$z$-plane}, the
integral (\ref{odd}) may be evaluated by contour integral
techniques. Let $C_\smV$ be a semicircle of radius $V$ centered at the
origin of the complex $z$-plane, with an orientation that starts at
$+V$ and ends at $-V$. We can traverse a closed
circuit by moving from $-V$ to $+V$ along the real axis, with
the circuit completed back to $-V$ by traversing $C_\smV$. The contour
integral around this closed circuit  vanishes since it contains no
interior singularities,
\begin{eqnarray}
  0
  &=&
  \oint dz \, 
  \frac{z}{k^2 + \kappa_e^2 + F_{\!\smI}(z)} 
\nonumber\\  
&=&
  \int_{-V}^V \! dv\, \frac{v}{k^2 + \kappa_e^2 + F_{\!\smI}(v)} 
  +
  \int_{C_\smV} \! dz \, \frac{z}{k^2 + \kappa_e^2 +
  F_{\!\smI}(z)} \,.
\end{eqnarray} 
Hence the integral (\ref{odd}) is equal to the negative of the
integral over the semicircle $C_\smV$ starting at $+V$ and ending at
$-V$. We can now take the limit $V \to \infty$ and use the asymptotic
form (\ref{zinf}) for $F_{\!\smI}$ along $C_\smV$. Since
\begin{equation}
\int_{C_\smV} dz \, z = i\, V^2\!\! \int_0^\pi d\theta \,
                          e^{2i\theta} = 0 \,,
\end{equation}
the leading term of the expansion of the denominator in
Eq.~(\ref{odd}) yields a vanishing contribution, and the only
non-vanishing term is given by first-order term with $F_\smI(z)$
replaced with its asymptotic form (\ref{zinf}):
\begin{eqnarray}
 \lim_{V\to \infty} i\int_{-V}^{+V} dv \, 
  \frac{v}{k^2 + \kappa_e^2 + F_{\!\smI}(v)} 
  &=&
  i \, \frac{1}{\left[ k^2 + \kappa_e^2 \right]^2} \, 
  \lim_{V\to\infty} \int_{C_\smV} dz \, z \, 
  \left[ - \frac{\omega_\smI^2}{z^2} \right]
\nonumber\\[5pt]
 &=&
 \frac{\pi}{\left[ k^2 + \kappa_e^2 \right]^2} \, \omega_\smI^2  \,.
\label{intlargev}
\end{eqnarray} 
Upon passing to hyper-spherical coordinates to perform the
$k$-integration, we now arrive at
\begin{eqnarray}
  \frac{\partial{\cal E}_{e \smI}^\smLT}{\partial t}
  &=&
  -  \frac{\beta_e \, e^2 \, m_e}{\hbar} \,\Big(T_e - T_\smI \Big) 
  \int \frac{d^{\nu-1}\, p_\smPerp}{(2\pi\hbar)^{\nu-1}} \, 
  \Big[f_e({\bf p}_\smPerp)\Big]^2 \,
  \exp\!\left\{ \beta_e \, \left[ \frac{p_\smPerp^{\,2}}{2m_e} - 
  \mu_e \right]\right\} 
\nonumber\\[5pt] && \hskip4cm   
  \omega_\smI^2 \, \frac{\Omega_{\nu-1}} {(2\pi)^\nu} \, 
  \int_0^\infty k^{\nu-1} d k \, k \, 
   \frac{1}{\left[ k^2 + \kappa_e^2 \right]^2} \,.
\label{dedt<<}
\end{eqnarray}
Changing variables by $k =t^{1/2} \kappa_e$ places the
$k$-integration in the form of a standard representation of the Euler
Beta function~\cite{abst},
and so we have
\begin{eqnarray}
  \int_0^\infty  \! d k \, 
  \frac{k^\nu}{\left[ k^2 + \kappa_e^2 \right]^2} =
  \frac{1}{2} \, \kappa_e^{\nu -3} \, 
  \frac{ \Gamma\left( \frac{\nu+1}{2} \right) \, 
  \Gamma\left( \frac{3-\nu}{2} \right) }{\Gamma(2)} \,.
\end{eqnarray}
Finally, we are now able to express the $\nu<3$ form of the
electron-ion energy exchange as
\begin{eqnarray}
  \frac{\partial{\cal E}_{e \smI}^\smLT}{\partial t}
  &=&
  -\frac{\beta_e\,e^2 \, m_e}{2\hbar}~
  \omega_\smI^2 \, \frac{\Omega_{\nu-1}} {(2\pi)^3} \, 
  \left( \frac{\kappa_e}{2\pi} \right)^{\nu -3} \, 
  \Gamma\left( \frac{\nu+1}{2} \right) \, 
  \Gamma\left( \frac{3-\nu}{2} \right) \Big(T_e - T_\smI \Big)
\nonumber\\[5pt]
  && \hskip3.5cm 
  \int \frac{d^{\nu-1} \,p_\smPerp}{(2\pi\hbar)^{\nu-1}} \, 
  \Big[f_e({\bf p}_\smPerp)\Big]^2 \, 
  \exp\!\left\{ \beta_e \, 
  \left[ \frac{p_\smPerp^{\,2}}{2m_e} - \mu_e \right]\right\} \ .
\label{dedt<<<}
\end{eqnarray}

\section{Adding the Rates}

The sum of the singular part (\ref{donees}) for the $\nu > 3$
contribution to the electron-ion energy exchange rate and the
$\nu < 3$ part (\ref{dedt<<<}) that we have just computed is
\begin{eqnarray}
  \frac{\partial{\cal E}_{e \smI}^{\smGT \, \smS}}{\partial t}+   
  \frac{\partial{\cal E}_{e \smI}^\smLT}{\partial t} 
  &=&
  -\frac{\beta_e \,e^2 \, m_e}{2\hbar}~ \omega_\smI^2 \,  
  \frac{\Omega_{\nu-1}}{(2\pi)^3} \,
  \Big( T_e - T_\smI \Big)
  \int \frac{d^{\nu-1} p_\smPerp}{(2\pi\hbar)^{\nu-1} } \,
  \Big[f_e({\bf p}_\smPerp) \Big]^2 \,
  \exp\!\left\{\beta_e \left[\frac{p_\smPerp^{\,2}}{2 m_e} -
  \mu_e \right]
  \right\} 
\nonumber\\[5pt]
&& \hskip1cm 
  \left( \frac{\pi\lambda_e^2}{4} \right)^{(3-\nu)/2}\,
  \left\{\Gamma\left(\frac{\nu -3}{2}\right) +
  \left( \frac{\kappa_e^2\lambda_e^2}{16\pi} \right)^{(\nu -3)/2} \, 
  \Gamma\left( \frac{\nu+1}{2} \right)\,\Gamma\left( \frac{3-\nu}{2} 
  \right) \right\} \,.
\label{almost}
\end{eqnarray}
As must be the case, the expression in the final curly braces
above is finite in the $\nu \to 3 $ limit.  To extract this limit,
we use
\begin{eqnarray}
\nonumber 
  \nu \to 3 \,: ~~ && 
\\[3pt] \nonumber
  && \hskip-1cm 
  \Gamma\left(\frac{\nu-3}{2}\right) \to \frac{2}{\nu-3} - \gamma  \ ,
  \hskip0.7cm 
  \Gamma\left(\frac{3-\nu}{2}\right) \to \frac{2}{3-\nu} - \gamma  \ ,
  \hskip0.7cm 
  \Gamma\left(\frac{\nu+1}{2}\right) \to 1 - (1 - \gamma) \,
  \frac{3-\nu}{2} 
\\
\end{eqnarray}
to evaluate the $\nu \to 3$ limit of the last line in
Eq.~(\ref{almost}):
\begin{eqnarray}
  \left[ \frac{2}{\nu-3} - \gamma \right] +
  \left( \frac{\kappa_e^2\lambda_e^2}{16\pi} \right)^{(\nu -3)/2} 
  \left[ \frac{2}{3-\nu} - 1 \right]
  \to 
  \ln\!\left\{ \frac{16\pi} {\kappa_e^2\lambda_e^2} \right\}-\gamma-1 \,.
\end{eqnarray}
Since this last factor is finite in the $\nu \to 3$ limit, we 
may now take the $\nu \to 3$ limit of all the other quantities in
Eq.~(\ref{almost}).  The integral over the perpendicular momenta
in this limit was evaluated previously in Eq.~(\ref{2d}), and so
we now have
\begin{eqnarray}
&&  
  \frac{\partial{\cal E}_{e \smI}^{\smGT \, \smS}}{\partial t} +   
  \frac{\partial{\cal E}_{e \smI}^\smLT}{\partial t} 
  =
  -\frac {\beta_e\, e^2 \, m_e}{2\hbar} \, 
  \frac{\omega_\smI^2 }{2\pi^2}\,\frac{1}{\lambda_e^2} \,
  \frac{1}{ \exp\!\left\{ - \beta_e \, \mu_e \right\}  +  1 } 
  \left[ \ln\!\left\{ \frac{16\pi} {\kappa_e^2\lambda_e^2} \right\}
  -\gamma - 1  \right] \Big( T_e - T_\smI \Big)\,.
\label{slim}
\end{eqnarray}
To this we must add the remaining finite part (\ref{finite}) of the
$\nu > 3$ contribution, namely
\begin{eqnarray}
  \frac{\partial{\cal E}_{e \smI}^{\smGT \, \smR}}{\partial t}
  &=& 
  -\frac{\beta_e\,e^2 m_e}{2\hbar}\, 
  \frac{\omega_\smI^2 }{2\pi^2}\, 
  \frac{1 }{\lambda_e^2} \,\Big( T_e - T_\smI \Big)
  \int_0^\infty dx \, \ln x
  \left\{ \frac{ \exp\!\left\{ x - \beta_e \, \mu_e  \right\} }
  {\left[ \exp\!\left\{ x - \beta_e \, \mu_e  \right\}  +  1 \right]^2 } 
  - \frac{ \exp\!\left\{- x \right\}}
  { \exp\!\left\{ - \beta_e \, \mu_e \right\}  +  1 } \right\} \,.
\label{finitee}
\end{eqnarray}
Recalling that we have defined [Eq.~(\ref{dedtei})]
\begin{eqnarray}
  \frac{d{\cal E}_{e \smI}}{dt}
  =   
  -\,{\cal C}_{e \smI} \Big( T_e - T_\smI \Big) \,,
\label{dedtEI}
\end{eqnarray}
we have now calculated the rate coefficient to leading order in the
plasma coupling and to all orders in the electron fugacity $z_e =
e^{\beta_e \mu_e}$\,,
\begin{eqnarray}
  {\cal C}_{e \smI} &=& \frac{\beta_e \, e^2 \, m_e}{2\hbar} \, 
  \frac{\omega_\smI^2 }{\pi^2} \,
  \frac{1}{\lambda_e^2} \, \Bigg\{
  \frac{1}{ \exp\!\left\{ - \beta_e \, \mu_e \right\}  +  1 } 
  \,\frac{1}{2}
  \left[ \ln\!\left\{ \frac{16\pi} {\kappa_e^2\lambda_e^2} \right\}
  -\gamma-1  \right] 
\nonumber\\[5pt]
&&
  + 
  \frac{1}{2}\int_0^\infty \!\! dx \, \ln x
  \left[ \frac{ \exp\!\left\{ x - \beta_e \, \mu_e  \right\} }
  {\left[ \exp\!\left\{ x - \beta_e \, \mu_e  \right\}  +  1 \right]^2 } 
  - \frac{ \exp\!\left\{- x \right\}}
  { \exp\!\left\{ - \beta_e \, \mu_e \right\}  +  1 } \right] \Bigg\} \,.
\label{donedeal}
\end{eqnarray}
By expanding the denominators, it is easy to check that
\begin{eqnarray}
  \int_0^\infty dx \, \ln x
  \left[ \frac{ \exp\!\left\{ x - \beta_e \, \mu_e  \right\} }
  {\left[ \exp\!\left\{ x - \beta_e \, \mu_e  \right\}  +  1 \right]^2 } 
  -\frac{ \exp\!\left\{- x \right\}}
  {\exp\!\left\{ - \beta_e \, \mu_e \right\}  +  1 } \right]  
  &=&  
  \sum_{l=1}^\infty (-1)^{l+1} \, \ln\{l+1\} \, e^{(l+1)\beta_e\mu_e}\,,
\nonumber\\
&&
\end{eqnarray}
which is an expansion in powers of the electron fugacity
$z_e = e^{\beta_e \mu_e}$. 

To place this result in a form that is easily compared to that of
BPS~\cite{bps}, we use the definition (\ref{deflambd}) of the 
thermal wavelength and a slight manipulation to write 
\begin{eqnarray}
  {\cal C}_{e \smI} 
  &=& 
  \frac{\omega_\smI^2}{2\pi}\,\sqrt{\frac{\beta_e m_e}{2\pi}} \,
  \left( \frac{2 \beta_e e^2}{\lambda_e^3} \, e^{\beta_e\mu_e}
  \right) \,  \Bigg\{
  \frac{1}{ \exp\!\left\{ \beta_e \, \mu_e \right\}  + 1}\, \frac{1}{2}\,
  \left[ \ln\!\left\{ \frac{16\pi}{\kappa_e^2\lambda_e^2} \right\}
  -\gamma - 1  \right] 
\nonumber\\[5pt]
&& \hskip3cm 
  + \, \frac{1}{2}
  \sum_{l=1}^\infty (-1)^{l+1} \, \ln\{l+1\} \, e^{l\beta_e\mu_e}\Bigg\} \,.
\label{donerdeal}
\end{eqnarray}
In the dilute limit in which Maxwell-Boltzmann statistics apply, the
fugacity $\exp\{\beta_e\mu_e\}$ is very small. The number density
approximation (\ref{napprox}) gives
\begin{eqnarray}
  \frac{2}{\lambda_e^3} \, e^{\beta_e \mu_e} &\simeq& n_e 
  \left[ 1 + \frac{1}{2^{3/2}} \, e^{\beta_e \mu_e} \right] \,,
\end{eqnarray}
and we see that keeping the first correction in the fugacity 
yields
\begin{eqnarray}
  {\cal C}_{e \smI} \simeq \frac{\omega_\smI^2}{2\pi} \,
  \sqrt{\frac{\beta_e m_e}{2\pi}} \, (\beta_e e^2 n_e) \, \left\{
  \left[ 1 - \left( 1 - \frac{1}{2^{3/2}} \right) \,
  e^{\beta_e\mu_e} \right] 
  \frac{1}{2} \left[ \ln\!\left\{  \frac{16\pi}{\kappa_e^2\lambda_e^2} \right\}
  - \gamma - 1  \right]  + \frac{1}{2} \,e^{\beta_e\mu_e} \, \ln 2 \right\} \,.
\label{agree}
\end{eqnarray}
Again remembering the fugacity approximation (\ref{kappae}),
which we repeat here,
\begin{eqnarray}
\nonumber\\
  \kappa_e^2 
  &\simeq& 
  \beta_e \, e^2 \, n_e \,  \left[
  1 - \frac{1}{2^{3/2}} \, e^{\beta_e \mu_e} \right] \,,
\end{eqnarray}
and the definitions
\begin{eqnarray}
  \lambda_e^2 = \frac{ 2\pi \, \hbar^2  \,\beta_e}{m_e} 
  \hskip1cm \text{and}\hskip1cm 
  \omega^2_e = \frac{e^2 n_e}{m_e} \,,
\end{eqnarray}
we find that
\begin{eqnarray}
  \frac{16\pi}{\kappa_e^2\lambda_e^2} 
  \simeq 
  \frac{8 T_e^2}{\hbar^2 \omega_e^2} \, \left[1 + \frac{1}{2^{3/2}}\, 
  e^{\beta_e \mu_e} \right] \,,
\end{eqnarray}
and thus
\begin{eqnarray}
  {\cal C}_{e \smI} 
  &\simeq& 
  \frac{\omega_\smI^2}{2\pi}\,\sqrt{\frac{\beta_e m_e}{2\pi}}\, 
  (\beta_e\, e^2 n_e) \, \Bigg\{\left[ 1 - \left(1 - 
  \frac{1}{2^{3/2}} \right)\,e^{\beta_e\mu_e} \right] 
  \frac{1}{2}\left[ \ln\!\left\{  \frac{8 T_e^2}{\hbar^2 \omega_e^2} \right\}
  -\gamma - 1  \right] 
\nonumber\\[5pt]
&& \hskip5.5cm 
  +\, e^{\beta_e\mu_e} \, \left[\frac{1}{2}\,\ln2 + 
  \frac{1}{2^{5/2}}\right]\Bigg\} \,.
\label{agreed}
\end{eqnarray}
\end{widetext}
We may use $e^{\beta_e \mu_e} \simeq \lambda_e^3 n_e/2$ inside the
curly braces of (\ref{agreed}). 
It is easy to confirm that with the neglect of the fugacity corrections,
this is in agreement with Eq.~(12.12) of BPS~\cite{bps} after that
equation is corrected as mentioned in the Introduction.

\begin{acknowledgments}
  We would like to thank George E. Cragg for reading the manuscript.
\end{acknowledgments}

\appendix
\begin{widetext} 

\section{\label{sec:difun} The Dielectric Function}
\onecolumngrid

In Section~\ref{sec:LBE}, the calculation of the rate in $\nu<3$ using
the Lenard-Balescu equation made extensive use of the plasma
dielectric function and its various properties.  The classical
dielectric function for a collisionless plasma is discussed in
Ref.~\cite{Lifs} for example, and the form of the result that we shall
use reads
\begin{eqnarray}
  \epsilon( k ,\omega) 
  =
  1 + {\sum}_b \, 
  \frac{e_b^2}{k^2} 
  \int\!\! \frac{d^\nu p_b}{(2\pi\hbar)^\nu}\, 
  \frac{1}{\omega - {\bf k} \!\cdot\! 
  {\bf v}_b + i \eta}\, {\bf k} \cdot 
  \frac{\partial}{\partial {\bf p}_b} f_b({\bf p}_b)  ,
\label{epsilonDef}
\\\nonumber
\end{eqnarray}
with the prescription $ \eta \to 0^+ $ defining the correct retarded
response. The degenerate electrons are described by the thermal
Fermi-Dirac distribution (\ref{defFeA}), so 
\begin{eqnarray}
  {\bf k} \cdot
  \frac{\partial}{\partial {\bf p}_e} f_e({\bf p}_e)
  &=&
  -\beta_e \, {\bf k} \cdot{\bf v}_e \,
  \frac{e^{\beta_e(E_e-\mu_e)}}{[e^{\beta_e(E_e -\mu_e)}+1]^2} 
\nonumber\\[5pt]
  &=&   -\beta_e \, {\bf k} \cdot {\bf v}_e \,
  f_e({\bf p}_e)\left[1 - f_e({\bf p}_e) \right] \,.
\label{eFD}
\\\nonumber
\end{eqnarray}
On the other hand, the ions are described by the Maxwell-Boltzmann
distribution (\ref{defFiA}), which is simply the large chemical
potential limit $-\beta\mu \gg 1$ of the Fermi-Dirac distribution.  In
this limit the Pauli blocking term is removed, $[1 - f({\bf p})\,] 
\to 1$, and so
\begin{eqnarray}
  {\bf k} \cdot
  \frac{\partial}{\partial {\bf p}_i} f_i({\bf p}_i)
  =
  -\beta_i \, {\bf k} \cdot{\bf v}_i \,  f_i({\bf p}_i) \,.
\label{MB}
\\\nonumber
\end{eqnarray}
For the real plasma considered in the text, the ions equilibrate to a
common temperature $T_\smI=1/\beta_\smI$; however, for the purposes of
this Appendix, we shall take each ion species $i$ to have an
individual inverse temperature $\beta_i$.  For degenerate electrons
and Maxwell-Boltzmann ions, the dielectric function (\ref{epsilonDef})
may therefore be expressed as
\begin{eqnarray}
  \epsilon( k ,\omega) &=& 1 -  \frac{\beta _e \, e^2}{k^2} \cdot
  2 \int \frac{d^\nu p_e}{(2\pi\hbar)^\nu}\, 
  \frac{ {\bf k} \cdot {\bf v}_e}{\omega - {\bf k} \!\cdot\! 
  {\bf v}_e + i \eta} \,  
  f_e({\bf p}_e)\left[1 - f_e({\bf p}_e) \right] 
\nonumber\\[5pt]
&& \qquad
 - {\sum}_i \, 
  \frac{\beta_i \, e_i^2}{k^2} 
  \int \frac{d^\nu p_i}{(2\pi\hbar)^\nu}\, 
  \frac{{\bf k} \cdot {\bf v}_i}{\omega - {\bf k} \!\cdot\! 
  {\bf v}_i + i \eta} \,  f_i({\bf p}_i) \,.
\\\nonumber
\label{epsAlg}
\end{eqnarray}
\end{widetext}

\noindent
The factor of two in the electron contribution arises from 
a sum over the two spin components of the electron.

The dielectric function in the Lenard-Balescu equation has the
functional form $\epsilon(k, {\bf v} \cdot {\bf k})$, with the speed
$|{\bf v}|$ much less than the electron thermal velocity.  Hence, in
the electron contribution to the dielectric function, the magnitude of
$ \omega = {\bf k} \cdot {\bf v}$ is much less than the typical
magnitude of ${\bf k} \cdot {\bf v}_e$, and we can use the $\omega \to 
0$ limit in which
\begin{eqnarray}
 && -\frac{\beta _e  e^2}{k^2} \cdot
  2 \int \frac{d^\nu p_e}{(2\pi\hbar)^\nu} 
  \frac{ {\bf k} \cdot {\bf v}_e}{\omega - {\bf k} \!\cdot\! 
  {\bf v}_e + i \eta}  
  f_e({\bf p}_e)\left[1 - f_e({\bf p}_e) \right]
\nonumber\\[5pt]
 && \hskip2cm \to 
  \frac{\kappa_e^2}{k^2} \,,
\label{kelim}
\end{eqnarray}
where 
\begin{eqnarray}
  \kappa_e^2 = 2\, \beta _e e^2
  \int \frac{d^\nu p_e}{(2\pi\hbar)^\nu}\, 
  f_e({\bf p}_e) \left[1 - f_e({\bf p}_e) \right] 
\label{kefe}
\end{eqnarray}
defines the electron contribution to the squared Debye wave number,
including the effects of Fermi-Dirac statistics which are explicitly
exhibited by the Pauli blocking factor $ \left[1 - f_e({\bf p}_e)
\right]$. From the form (\ref{defFeA}) of the thermal Fermi-Dirac
distribution $f_e({\bf p_e})$ and the definition (\ref{fbnorm}) of the
number density, we see that
\begin{eqnarray}
  \kappa_e^2 
  &=&   
  e^2 \beta_e \, \frac{\partial n_e}{\partial (\beta_e \mu_e)} \,.
\label{kefluc}
\end{eqnarray}
Remembering the structure of the grand canonical ensemble, the
derivative that appears here is the thermal average of the fluctuations
about the mean particle number.  In the limit of Maxwell-Boltzmann
statistics, the derivative simply reproduces the particle number
density, corresponding to the fact that classical statistics have a
Poisson distribution.  Multiplying Eq.~(\ref{epsAlg}) by $k^2$ and
using Eq.~(\ref{kelim}) for the electron contribution, we can write
\begin{eqnarray}
  k^2 \,   \epsilon( k ,\omega) = k^2 + \kappa_e^2 + 
  F_{\!\smI}(\omega/k) \,,
\label{Fdef}
\end{eqnarray}
where we have defined the function
\begin{eqnarray}
  F_{\!\smI}(v) &=& - {\sum}_i \, \beta_i \, e_i^2 \,
  \int \frac{d^\nu p_i}{(2\pi\hbar)^\nu}\, 
  \frac{\hat{\bf k} \cdot {\bf v}_i}{ v  - \hat{\bf k} \!\cdot\! 
  {\bf v}_i + i \eta} \,  f_i({\bf p}_i) \,.
\nonumber\\
&&
\label{Fform}
\end{eqnarray}
This is almost same function $F$ defined in Ref.~\cite{bps}, except
that here we have handled the electron contribution separately.

Superficially it appears that $F_{\!\smI}$ contains wave-vector
dependence through the terms $\hat{\bf k}\cdot {\bf v}_i$ in the
integrand of (\ref{Fform}); however, since we are integrating over all
values of ${\bf v}_i$, the wave-vector direction $\hat{\bf k}$ cancels
in $F_{\!\smI}$.  As our notation suggests, $F_{\!\smI}(v)$ is indeed
only a function of \hbox{$v = |{\bf v}|$}. Furthermore, because of the
\hbox{$i\eta$}-term with $\eta>0$ in the denominator, the function
$F_{\!\smI}$ is analytic in the upper complex $v$-plane.

In evaluating the integral (\ref{intlargev}) in the text, we require
the large-$v$ behavior of (\ref{Fform}). Since the numerator of
the integrand in Eq.~(\ref{Fform}) is odd, we can expand the 
denominator to find the leading $v$-behavior 
\begin{widetext}
\begin{eqnarray}
  |v| \to \infty \,: \hskip1cm 
  F_{\!\smI}(v) 
  \to 
  -\sum_i \beta_i\,e_i^2\,\int \frac{d^\nu p_i}{(2\pi\hbar)^\nu} \, 
  \frac{(\hat{\bf k} \cdot {\bf v}_i)^2}{v^2} \, f_i({\bf p}_i) 
  = 
  - \frac{\omega_\smI^2}{v^2} + {\cal O}(v^{-4})\,,
\label{Fvlim}
\end{eqnarray}
where
\begin{eqnarray}
  \omega_\smI^2 = {\sum}_i \, \frac{e_i^2 \, n_i}{m_i}
\end{eqnarray}
is the sum of the squared ion plasma frequencies. 
To obtain this result, the Gaussian
integral in Eq.~(\ref{Fvlim}) may be calculated directly, or more
elegantly, one may use Eq.~(\ref{MB}) to replace $(\hat{\bf k}\cdot
{\bf v}_i) f_i$ with a derivative of $f_i$, after which a partial
integration yields Eq.~(\ref{Fvlim}).

Finally, we derive a dispersion relation that will be quite useful in
evaluating Eq.~(\ref{dedt<}) in Section~\ref{sec:LBE}. Applying the
relation
\begin{eqnarray}
  {\rm Im} \, \frac{1}{x + i\eta} = - \pi \delta(x) \,
\end{eqnarray}
for $\eta \to 0^+$ in Eq.~(\ref{Fform}) allows us to express the 
imaginary part of $F_\smI$ in the form
\begin{eqnarray}
  {\rm Im} \, F_{\!\smI}(v) = {\sum}_i \, 
  \beta_i \, e_i^2 \,
  \int \frac{d^\nu p_i}{(2\pi\hbar)^\nu} \,  f_i({\bf p}_i) \, 
  v \, \pi \, \delta( v  - \hat{\bf k} \!\cdot\!  {\bf v}_i ) \,.    
\end{eqnarray}
From this, we can find the imaginary part of the inverse of
the dielectric function:
\begin{eqnarray}
  {\rm Im} \, \frac{1}{k^2 \, \epsilon( k ,{\bf v}\cdot{\bf k})} 
  &=&
  -\frac{{\rm Im}\,k^2\,\epsilon(k ,{\bf v}\cdot{\bf k})} 
  {|k^2 \,\epsilon( k ,{\bf v}\cdot{\bf k})|^2} 
  = 
  -\frac{{\rm Im}\,F_{\!\smI}(v)}
  {|k^2 \, \epsilon( k ,{\bf v}\cdot{\bf k} )|^2}
\nonumber\\[5pt]
  &=& 
  - \frac{1}{|k^2 \, \epsilon( k ,{\bf v}\cdot{\bf k} )|^2} \,
 {\sum}_i \beta_i\, e_i^2
  \int \frac{d^\nu p_i}{(2\pi\hbar)^\nu} \,  f_i({\bf p}_i) \,
  v \, \pi \, \delta( v  - \hat{\bf k} \!\cdot\!  {\bf v}_i ) \,. 
\end{eqnarray}
Since the numerator in the integrand (\ref{Fform}) is odd, under
complex conjugation we have
\begin{eqnarray}
  F_{\!\smI}(-v) = F_{\!\smI}(+v)^* \,.
\label{refl}
\end{eqnarray}
Hence, using Eq's.~(\ref{Fdef}) and (\ref{refl}) can write 
\begin{eqnarray}
&& \frac{\pi v}{|k^2 \, \epsilon( k ,{\bf v}\cdot{\bf k} )|^2} \,
 {\sum}_i \, 
  \beta_i \, e_i^2 \,
  \int \frac{d^\nu p_i}{(2\pi\hbar)^\nu} \,  f_i({\bf p}_i) \,
  \delta(\hat{\bf k} \!\cdot\!  {\bf v}_i - v) \,. 
\nonumber\\[5pt]
&& \qquad\qquad\qquad\qquad =
- \frac{1}{2i} \, \left\{ \frac{1}{k^2 + \kappa_e^2 + F_{\!\smI}(v)} 
   -  \frac{1}{k^2 + \kappa_e^2 + F_{\!\smI}(-v)} \right\} \,.
\label{wonder}
\end{eqnarray}
\end{widetext}

\section{\label{sec:scatcorr} Scattering Corrections}

The electron-ion energy exchange rate computed in Section~12 of 
BPS~\cite{bps}
was performed under quite general conditions, with no restriction on
the masses, number densities, or temperatures of the plasma
components, except that the plasma be fully ionized, non-degenerate,
and weakly to moderately coupled (all mild restrictions in a hot,
low-$Z$ plasma). Because of its generality, this result, which we
shall present momentarily, is rather complicated. However, for most
practical calculations, we can work in the high-temperature
extreme-quantum limit and take advantage of the small electron-to-ion
mass ratio. Under these conditions, we can use the Born approximation
for the two-body scattering amplitude, and the rate coefficient
collapses to the simple expression, 
\begin{eqnarray}
  {\cal C}_{e \smI} 
  &=& 
  \frac{\omega_\smI^2}{2\pi}\,
  \sqrt{\frac{\beta_e m_e}{2\pi}}\, ( \beta_e e^2 n_e )
  \frac{1}{2}\left[ \ln\!\left\{  \frac{8 T_e^2}
  {\hbar^2 \omega_e^2} \right\} - \gamma - 1  \right] \,,
\nonumber\\
&&
\label{CeIsimple}
\end{eqnarray}
as previously quoted in ~(\ref{nondegenrate}). The purpose
of this Appendix is to find the subleading quantum correction to
Eq.~(\ref{CeIsimple}). 

As noted in BPS, this subleading correction is of order 
\begin{eqnarray}
  \eta_{ei}^2 \sim 27\,{\rm eV}/T_e \ , 
\end{eqnarray}
which, for most applications that we have in
mind, is quite small. However, the leading electron degeneracy effects
are of order $z_e \sim n_e\, a_0^3\, (27\,{\rm eV}/T_e)^{3/2}$, which
can be comparable in size to the subleading quantum correction. Both
corrections are small compared to the leading-order
contribution~(\ref{CeIsimple}). Therefore, in this Appendix, we
can work in the non-degenerate limit, since degeneracy effects on top
of the subleading quantum effects are smaller still.  Consequently,
our starting point will be the non-degenerate, but otherwise rather
general, expression for the rate derived in Section~12 of BPS. We
shall simplify this rate in favor of the more realistic case of light
electrons and heavy ions, exhibiting this result in
Eq.~(\ref{Ceialleta}), an expression that is valid to all orders in
the quantum parameter $\eta_{ei}$. Although this expression is clear
and compact, for the purposes of this paper, however, we only need the
subleading $\eta_{ei}^2$ term. This subleading correction is displayed
in Eq.~(\ref{firstcl}).

The strength of the quantum effects associated with the scattering of 
two plasma species $a$ and $b$ is characterized by the dimensionless 
parameter
\begin{eqnarray}
  \bar\eta_{ab} 
  &=& 
  \frac{e_a e_b}{4\pi \hbar V_{ab}}  \ ,
\label{etaab}
\end{eqnarray}
where the square of the thermal velocity in this expression is defined
by
\begin{eqnarray}
  V_{ab}^2 
  =
  \frac{1}{\beta_a m_a} + \frac{1}{\beta_b m_b} \ .
\label{Vab}
\end{eqnarray}
The extreme quantum limit, where formally $\hbar \to \infty$, 
is given by $\bar\eta_{ab}\to 0$; while the extreme classical limit,
where formally $\hbar \to 0$,  is given by
$\bar\eta_{ab}\to\infty$.  The former case is equivalent to the Born
approximation.  In Section~12 of BPS, the energy exchange
rate from an arbitrary plasma species $a$ to another species $b$,
\begin{eqnarray}
  \frac{d{\cal E}_{ab}}{dt}
  =   -\, {\cal C}_{ab} \left( T_a - T_b \right) \ ,
\label{dedtab}
\end{eqnarray}
was computed to all orders in the two-body quantum-scattering
parameter $\bar \eta_{ab}$. It was found that the rate coefficient can
be written as a sum of three terms, which, in the notation of BPS,
reads
\begin{eqnarray}
  {\cal C}_{ab}
  =
  {\cal C}^\smLT_{ab,\smR} 
  +
  \Big({\cal C}^\smC_{ab,\smS} +
  {\cal C}^\smDelQ_{ab} \Big) \ ,
\end{eqnarray}
where the last two terms have been grouped together for later
convenience. These three terms are given by Eq's.~(12.31), (12.25), and
(12.50) respectively in BPS:
\begin{widetext}
\begin{eqnarray}
  {\cal C}^\smLT_{ab,\smR} 
  &\!=\!&
  \frac{\kappa_a^2\, \kappa_b^2}{2\pi}
  \left(\frac{\beta_a m_a}{2\pi} \right)^{\!\! 1/2} 
  \left(\frac{\beta_b m_b}{2\pi} \right)^{\!\! 1/2} \!\!
  \int_{-\infty}^{\infty} \!\! dv \, v^2 
  e^{- \frac{1}{2}( \beta_a m_a + \beta_b m_b) v^2 }  
  \frac{i}{2 \pi} \,\frac{F(v)}{\rho_\text{total}(v)}\, 
  \ln\! \left\{ \frac{F(v)}{K^2}\right\} ,
\label{nunnn}
\end{eqnarray}
\begin{eqnarray}
  {\cal C}^\smC_{ab,\smS} 
  &\!=\!& 
  \!-{\kappa_a^2\, \kappa_b^2 } \, 
  \frac{ (\beta_a m_a \beta_b m_b)^{1/2}}{\left( \beta_a m_a + 
  \beta_b m_b \right)^{3/2} } \,\left( \frac{1}{2\pi} \right)^{\!\!3/2}\, 
  \left[\,\ln\!\left\{ \frac{e_a\,e_b}{4 \pi}\,
  \frac{K}{4 \, m_{ab} \, V^2_{ab}}\right\} 
  + 2 \gamma\,  \right] \ ,
\label{classicdone}
\end{eqnarray}
and 
\begin{eqnarray}
 {\cal C}^{\Delta Q}_{ab} 
  &\!=\!&  
  \!-\frac{1}{2} \, \kappa_a^2\, \kappa_b^2 \, 
  \frac{(\beta_a m_a \, \beta_b m_b)^{1/2}}{(\beta_a m_a \!+\! 
  \beta_b m_b)^{3/2}} 
  \left(\frac{1}{2\pi}\right)^{\!\!3/2} \!\!\!\!
  \int_0^\infty \!\!\! d \zeta\, e^{-\zeta/2} 
  \left[\,{\rm Re}\,
  \psi\!\left(\!1 + i\frac{\bar\eta_{ab}}{\zeta^{1/2}}\right) \!-\! 
  \ln\!\left\{ \frac{\bar\eta_{ab}}{\zeta^{1/2}}\right\}
  \right] \,.
\label{qqcorr}
\end{eqnarray}
\end{widetext}
Since we are using Maxwell-Boltzmann statistics throughout this section,
the Debye wave number of any plasma species, including
electrons, is here determined by $\kappa_b^2 = \beta_b\, e_b^2\, n_b$. 
The complex-valued function $F(v)$ is defined by
\begin{eqnarray}
  F(v) 
  = 
  \int_{-\infty}^\infty \! du \, 
  \frac{\rho_\text{total}(u)}{v - u + i\eta} \ ,
\end{eqnarray}
where $\rho_\text{total}(v)$ is the spectral weight,
\begin{eqnarray}
  \rho_\text{total}(v)
  &=&
  {\sum}_b\, \rho_b\!\left(v\right) \ ,
\end{eqnarray}
with
\begin{eqnarray}
  \rho_b(v) 
  = 
  \kappa_b^2\,\sqrt{\frac{\beta_b m_b}{2\pi}}\, v\,
  \exp\!\left\{-\frac{1}{2}\,\beta_b m_b\, v^2\right\} \ .
\label{rhototdef}
\end{eqnarray}
This is similar to the function $F_{\!\smI}$ introduced in the
previous Appendix, except here the sum extends over all plasma
species, including electrons.  Note the dependence in
Eq's.~(\ref{nunnn}) and~(\ref{classicdone}) on an unspecified parameter
$K$, with the only restriction being that $K$ has units of a
wave-number. This is an artifact of the calculational procedure, and
it was shown in BPS that the total rate ${\cal C}_{ab}$ is indeed
independence $K$, as this parameter cancels in the sum between
Eq's.~(\ref{nunnn}) and (\ref{classicdone}). As a matter of technical
convenience, we will henceforth set $K=\kappa_e$ throughout the rest of
this Appendix [the simplified result (\ref{smart}) only holds under
this condition].  Finally, we should note that the reduced mass is 
determined by $1/m_{ab}=1/m_a + 1/m_b$, and 
\begin{eqnarray}
  \psi(z) = \frac{1}{\Gamma(z)}\, \frac{d\Gamma(z)}{dz} 
\end{eqnarray}
is the logarithmic derivative of the gamma function.

Specializing to electrons and ions (in which $a\!=\!e$ and
$b\!=\!i$), we can employ Eq.~(\ref{rhototdef}) to write
Eq.~(\ref{nunnn}) in the form
\begin{widetext}
\begin{eqnarray}
  {\cal C}^\smLT_{e\smI,\smR} 
  &\!\! \equiv \!\!&
  {\sum}_i{\cal C}^\smLT_{e i,\smR} 
  \!=\!
  \frac{\kappa_e^2}{2\pi}
  \left(\frac{\beta_e m_e}{2\pi} \right)^{1/2} \!
  \int_{-\infty}^{\infty} \!\! dv \, v \,
  e^{- \frac{1}{2}\beta_e m_e v^2 }  
  \frac{i}{2 \pi} \,\frac{{\sum}_i \rho_i(v)}{\rho_\text{total}(v)}\, 
  F(v)\ln\! \left\{ \frac{F(v)}{\kappa_e^2}\right\} \ .
\label{nunnnion}
\end{eqnarray}
This expression greatly simplifies since the electron is so much
lighter than the ions. In virtually all practical applications,
the electron and ion temperatures are never excessively disparate,
and we can therefore impose the mild restriction 
\begin{eqnarray}
  \beta_e m_e \ll \beta_i m_i \ .
\label{lightme}
\end{eqnarray}
We will refer to the condition (\ref{lightme}) as the $m_e \to 0$
limit, and the ratio $\beta_e m_e/\beta_im_i$ can then be used as a
small dimensionless expansion parameter. For example, to leading order in this
parameter we find
\begin{eqnarray}
  \frac{{\sum}_i \rho_i(v)}{\rho_\text{total}(v)}
  =
  1 + {\cal O}\!\left(\frac{\beta_e m_e}{\beta_im_i}\right)^{\!1/2} \ ,
\end{eqnarray}
which allows us to express Eq.~(\ref{nunnnion}) as
\begin{eqnarray}
  {\cal C}^\smLT_{e\smI,\smR} 
  =
  \frac{\kappa_e^2}{2\pi}
  \left(\frac{\beta_e m_e}{2\pi} \right)^{\!1/2} 
  \frac{i}{2 \pi}\! \int_{-\infty}^{\infty} \!\! dv \,v\,F(v)\, 
  \ln\!\left\{ \frac{F(v)}{\kappa_e^2}\right\} 
  \Bigg[1 + {\cal O}\!\left(\frac{\beta_e m_e}{\beta_i m_i}
  \right)^{\!1/2}\Bigg]  .
\label{nunnnsmallme}
\end{eqnarray}
We have omitted the exponential in the integrand of
(\ref{nunnnion}), since the function $F(v)$ provides enough
convergence at large values of $v$ to allow the $m_e \to 0$ limit 
to be brought inside the integral. 
The analytic properties of $F(v)$ allow us to
perform the $v$ integral in Eq.~(\ref{nunnnsmallme}) using contour
integral techniques, in much the same manner as we did in the
discussion following Eq.~(\ref{odd}) in the text. The result is
Eq.~(12.44) of BPS, which reads
\begin{eqnarray}
  \beta_e m_e \ll \beta_i m_i \,: \hskip1cm 
  {\cal C}^\smLT_{e\smI,\smR} 
  &=&
  - \frac{1}{2}\,\frac{\kappa_e^2}{2\pi} \,
  \left( \frac{\beta_e m_e}{2\pi} \right)^{1/2} \, {\sum}_i\omega_i^2 \ ,
\label{smart}
\end{eqnarray}
a much simpler expression indeed. For the electrons and ions we are
considering, we can also drop the term of order $\beta_em_e/ \beta_i
m_i$ in Eq.~(\ref{etaab}), allowing us to express the quantum
parameter of Eq.~(\ref{Vab}) as
\begin{eqnarray}
  \bar\eta_{ei} 
  &=& 
  \frac{Z_i\,e^2}{4\pi \hbar} \, \sqrt{\frac{m_e}{T_e}} 
  = Z_i \,\bar\eta_e\ .
\label{etaeidef}
\end{eqnarray}
On occasion, we will express the quantum parameter in terms of the
binding energy of the hydrogen atom
\begin{eqnarray}
  \epsilon_\smH 
  =
  \frac{1}{2}\left(\frac{e^2}{4\pi}\right)^2 \frac{m_e}{\hbar^2}
  \simeq 
  13.606 \, {\rm eV} \ ,
\label{epsH}
\end{eqnarray}
so that
\begin{eqnarray}
  \bar\eta^2_{ei} 
  = 
  Z_i^2 \, \frac{2\, \epsilon_\smH}{T_e} \,.
\label{etaeiH}
\end{eqnarray}
We see that $\bar \eta_{ei} \ll 1$ when $T_e$ reaches the keV scale,
illustrating that quantum corrections are important at high
temperatures.  Finally, we can drop terms of order $\beta_em_e/
\beta_i m_i$ in the leading coefficients of Eq's.~(\ref{classicdone})
and (\ref{qqcorr}), thereby allowing us to write
\begin{eqnarray}
\nonumber
  \beta_e m_e \ll \beta_i m_i \,:
\hskip1cm  && 
\nonumber\\[-5pt]
\nonumber
  {\cal C}^\smC_{e \smI,\smS} +  {\cal C}^{\Delta Q}_{e \smI} 
  &=& 
  -
  \frac{\kappa_e^2}{2\pi}\, 
  \left(\frac{\beta_e m_e}{2\pi}\right)^{1/2} 
  {\sum}_i \omega_i^2
  \Bigg[\,\ln\!\left\{ \frac{Z_i\,e^2}{4\pi}\,
  \frac{\kappa_e}{4 T_e}\right\} + 2 \gamma\,
\\&&
  +\, \frac{1}{2}\,
  \int_0^\infty \!\! d\zeta\, e^{-\zeta/2} \!
  \left( {\rm Re}\,\psi\!\left( 1 + i\,
  \frac{Z_i\,\bar\eta_e}{\zeta^{1/2}}  \right) - 
  \ln\!\left\{ \frac{Z_i\,\bar\eta_e}{\zeta^{1/2}}\right\}\,\right)~
  \Bigg] \ ,
\label{csqmsmall}
\end{eqnarray}
Unlike Eq.~(\ref{smart}), which only holds for the sum over ion
components, the result (\ref{csqmsmall}) actually holds component by
component. Performing the $\zeta$ integral for the last term in
Eq.~(\ref{csqmsmall}) gives
\begin{eqnarray}
\nonumber
  \beta_e m_e \ll \beta_i m_i \,:
\hskip-0.7cm && 
\nonumber\\
\nonumber
  {\cal C}^\smC_{e \smI,\smS} \!+\!  {\cal C}^{\Delta Q}_{e \smI} 
  &=& 
  \frac{\kappa_e^2}{2\pi}\, 
  \left(\frac{\beta_e m_e}{2\pi}\right)^{1/2} \!\!
  {\sum}_i \frac{\omega_i^2}{2}
  \Bigg[\ln\!\left\{ \frac{8 m_e T_e}{\hbar^2 \kappa_e^2}\right\}\! 
  - 3 \gamma  \!-\!\!
  \int_0^\infty \!\! d \zeta\, e^{-\zeta/2} \,
  {\rm Re}\,\psi\!\left( 1 + i\,
  \frac{\bar\eta_{ei}}{\zeta^{1/2}}  \right)
  \Bigg] \ .
\\
\label{csqmsimple}
\end{eqnarray}
The rate ${\cal C}_{e\smI}$ is given by adding Eq's.~(\ref{smart}) and
(\ref{csqmsimple}), which can be written as
\begin{eqnarray}
\nonumber
  \beta_e m_e \ll \beta_i m_i \,: && 
\\ 
  && \hskip-3cm 
  {\cal C}_{e \smI} 
  = 
  \frac{\kappa_e^2}{2\pi}\, 
  \left(\frac{\beta_e m_e}{2\pi}\right)^{1/2} 
  \frac{1}{2}\,{\sum}_i\omega_i^2
  \Bigg[\ln\!\left\{ \frac{8 m_e T_e}{\hbar^2 \kappa_e^2}\right\} 
  -\gamma -1  - \Delta_i(\bar\eta_{ei})
  \Bigg] \ ,
\label{Ceialleta}
\end{eqnarray}
with
\begin{eqnarray}
  \Delta_i(\bar\eta_{ei})
   = 
  \int_0^\infty \!\! d \zeta\, e^{-\zeta/2} \,
  \Bigg[ {\rm Re}\,\psi\!\left( 1 + i\,
  \frac{\bar\eta_{ei}}{\zeta^{1/2}}\right) + \gamma
  \Bigg] \ .
\end{eqnarray}
This expression is accurate to leading and next-to-leading order in
the plasma coupling, and to all orders in $\bar\eta_{ei}$, with no
restriction on the temperature (apart from requiring the mild
constraint (\ref{lightme}) and that the plasma coupling be
small). 

With the aid of Eq.~(10.17) of BPS,
\begin{eqnarray}
  {\rm Re}\,\psi\left(1 + i\,\bar\eta\,\zeta^{-1/2}\right)
  + \gamma = \sum_{k=1}^\infty \, \frac{1}{k} \, 
  \frac{\bar\eta^2}{k^2 \, \zeta + \bar\eta^2} \,,
\end{eqnarray}
and writing 
\begin{eqnarray}
  \frac{1}{\zeta + \bar\eta^2/ k^2} 
  =
  \frac{d}{d\zeta}\,\ln\!\left\{ \zeta + 
\frac{\bar\eta^2}{k^2} \right\}  \,,
\end{eqnarray}
a partial integration now gives
\begin{eqnarray}
  \Delta_i(\bar\eta_{ei})
  =
  \bar\eta_{ei}^2 \, \sum_{k=1}^\infty \, \frac{1}{k^3}\, 
  \Bigg[  \ln\!\left\{\frac{k^2}{\bar\eta_{ei}^2} \right\} + 
  \frac{1}{2} \,  \int_0^\infty d\zeta  \, e^{- \zeta/2} \, 
  \ln\!\left\{ \zeta +  \frac{\bar\eta_{ei}^2}{k^2} \right\} 
  \Bigg] \,. 
\end{eqnarray}
In the latter form, we can easily extract the leading order
term $\bar\eta_{ei}$, since we can use the limit 
\begin{eqnarray}
\nonumber
  \bar\eta_{ei} \to 0 \,: \hskip5.5cm  && 
\\
  \frac{1}{2}\int_0^\infty d\zeta  \, e^{- \zeta/2} \, 
  \ln\!\left\{ \zeta +  \frac{\bar\eta_{ei}^2}{k^2} \right\} 
  &\to&
  \int_0^\infty d(\zeta/2)  \, e^{- \zeta/2} \, 
  \Big( \ln\!\left\{ \zeta /2 \right\} + \ln 2 \Big)
\\[5pt]
\nonumber
  &=&  -\gamma + \ln2 \,.
\end{eqnarray}
Using now
\begin{eqnarray}
  \zeta(3) 
  =  
  \sum_{k=1}^\infty \, \frac{1}{k^3}  
  = 
  1.20205 \cdots \,,
\end{eqnarray}
and
\begin{eqnarray}
  \zeta'(3) 
  =  
  -\sum_{k=1}^\infty \, \frac{1}{k^3} \, \ln k 
  = 
  -0.19812 \cdots \,,
\end{eqnarray}
we can express the subleading quantum correction as
\begin{eqnarray}
  \Delta_i(\bar\eta_{ei})
  \simeq 
  \bar\eta_{ei}^2 \left\{
  \zeta(3)\!\left[ \ln\!\left\{\frac{2}{\bar\eta_{ei}^2}\right\}
  - \gamma \right] -2\,\zeta'(3)
  \right\} \,. 
\end{eqnarray}
Using Eq.~(\ref{etaeiH}), we can express the rate (\ref{Ceialleta}) 
to leading and next-to-leading order in $\bar\eta_{ei}$ as
\begin{eqnarray}
  {\cal C}_{e \smI} 
  &=& 
  \frac{\omega_\smI^2}{2\pi}  \,
  \sqrt{\frac{\beta_e m_e}{2\pi}}~ \kappa_e^2 ~ \frac{1}{2}\left[
  \ln\!\left\{ \frac{8\,T_e^2}{\hbar^2\,\omega_e^2} \right\}
  -\gamma - 1\, \right]
\nonumber\\[5pt]
&& \quad
  -\,\frac{1}{2\pi}\, \sqrt{\frac{\beta_e m_e}{2\pi}} ~\kappa_e^2~~
  \frac{\epsilon_\smH}{T_e} {\sum}_i \, Z_i^2 \, \omega_i^2 \,
  \left[\zeta(3)\left(
  \ln\!\left\{\frac{T_e}{Z_i^2\,\epsilon_\smH} \right\}
  -\gamma \right) - 2\,\zeta'(3) \right]  \,. 
\label{firstcl}
\end{eqnarray}
The correction is of order $\epsilon_\smH/T_e$, which can be of
comparable size to the leading degeneracy correction.
$$
$$
\end{widetext}


\begin{thebibliography}{99}
\bibitem{born}
  The validity of using the Born approximation for the scattering
  amplitude is discussed in more detail in Ref.~\cite{bps}, 
  particularly in the work leading to its Eq.~(12.54).
\end{thebibliography}
\end{document}